\documentclass{IEEEtran4PSCC}
\makeatletter
\renewcommand{\thanksto}[1]{%
  \par\medskip
  \noindent #1\par
}
\makeatother
\ifCLASSINFOpdf
\usepackage[pdftex]{graphicx}
\usepackage[cmex10]{amsmath}

\makeatletter
\let\old@ps@headings\ps@headings
\let\old@ps@IEEEtitlepagestyle\ps@IEEEtitlepagestyle
\def\psccfooter#1{%
    \def\ps@headings{%
        \old@ps@headings%
        \def\@oddfoot{\strut\hfill#1\hfill\strut}%
        \def\@evenfoot{\strut\hfill#1\hfill\strut}%
    }%
    \def\ps@IEEEtitlepagestyle{%
        \old@ps@IEEEtitlepagestyle%
        \def\@oddfoot{\strut\hfill#1\hfill\strut}%
        \def\@evenfoot{\strut\hfill#1\hfill\strut}%
    }%
    \ps@headings%
}
\makeatother
\psccfooter{
        \parbox{\textwidth}{\hrulefill \\ \small{24th Power Systems Computation Conference} \hfill \begin{minipage}{0.2\textwidth}\centering \vspace*{4pt} \includegraphics[scale=0.06]{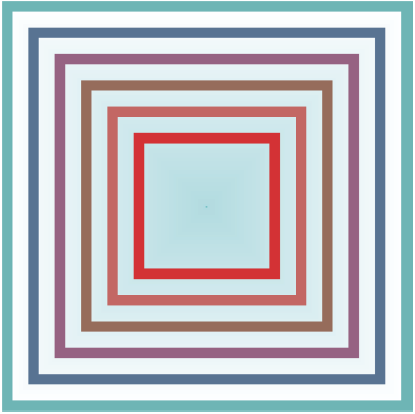}\\\small{PSCC 2026} \end{minipage} \hfill \small{Limassol, Cyprus --- June 8-12, 2026}}}

\usepackage{amsmath,amssymb}
\usepackage{booktabs}
\usepackage{caption}
\usepackage{microtype}
\usepackage{siunitx}
\usepackage{stfloats}
\usepackage{indentfirst}

\usepackage{subcaption} 
\IEEEoverridecommandlockouts
\usepackage{cite}
\usepackage{amsmath,amssymb,amsfonts}
\usepackage{bigdelim}
\usepackage{multirow}
\usepackage{graphicx}
\usepackage{subcaption}
\usepackage[dvipsnames]{xcolor}
\definecolor{ACgreen}{HTML}{2E7D32}
\usepackage{tikz}
\usetikzlibrary{decorations.pathreplacing,calc}

\usepackage{tikz}
\usetikzlibrary{decorations.pathreplacing}

\usepackage{microtype} 
\usepackage{enumitem}  
\usepackage{graphicx} 
\usepackage{xcolor}
\definecolor{darkgreen}{rgb}{0.0,0.5,0.0}
\usepackage{graphicx,adjustbox}
\usepackage{algpseudocode}
\usepackage{algorithm}
\usepackage{graphicx}
\usepackage{textcomp}
\usepackage{xcolor}
\usepackage{float}
\usepackage{mathtools}
\usepackage{booktabs}
\usepackage{multirow}
\usepackage{comment}
\usepackage{enumitem}  
\usepackage{balance}
\usepackage{hyperref}
\hypersetup{
    colorlinks=true,
    linkcolor=blue,
    filecolor=blue,      
    urlcolor=blue,
    citecolor=blue,
    pdftitle={TowardsACFeasibility}}
    
\usepackage{caption}
\usepackage{subcaption}
\usepackage{float}
\usepackage{calc} 
\newlength{\forallw}
\makeatother
\usepackage{bm}  
\usepackage{enumitem}
\usepackage[compact]{titlesec}
\titlespacing*{\section}{0pt}{*1.4}{*0.85} 

\def\BibTeX{{\rm B\kern-.05em{\sc i\kern-.025em b}\kern-.08em
    T\kern-.1667em\lower.7ex\hbox{E}\kern-.125emX}}

\begin{document}

\bstctlcite{BSTcontrol}

 \title{Towards AC Feasibility of DCOPF Dispatch

\vspace{-0.5em}
}

\author{%
\IEEEauthorblockN{Michael A. Boateng\IEEEauthorrefmark{1},
Russell Bent\IEEEauthorrefmark{2},
Sidhant Misra\IEEEauthorrefmark{2},
Parikshit Pareek\IEEEauthorrefmark{3}, 
Pascal Van Hentenryck\IEEEauthorrefmark{1},
Daniel Molzahn\IEEEauthorrefmark{1}}
\IEEEauthorblockA{\IEEEauthorrefmark{1} School of Electrical and Computer Engineering, Georgia Institute of Technology
\\Atlanta, Georgia, USA, \{mboateng6, pvh, molzahn\}@gatech.edu}
\IEEEauthorblockA{\IEEEauthorrefmark{2} T5: Applied Mathematics and Plasma Physics, Los Alamos National Laboratory
\\Los Alamos, New Mexico, USA, \{rbent, sidhant\}@lanl.gov}
\IEEEauthorblockA{\IEEEauthorrefmark{3} 
Department of Electrical Engineering, Indian Institute of Technology Roorkee, India, pareek@ee.iitr.ac.in}}

\vspace{-2em}

\maketitle
\begin{abstract} 
DC Optimal Power Flow (DCOPF) is widely utilized in power system operations due to its simplicity and computational efficiency. However, its lossless, reactive power–agnostic model often yields dispatches that are infeasible under practical operating scenarios such as the nonlinear AC power flow (ACPF) equations. While theoretical analysis demonstrates that DCOPF solutions are inherently AC-infeasible, their widespread industry adoption suggests substantial practical utility. This paper develops a unified DCOPF $\!\rightarrow\!$ ACPF pipeline 
to recover AC feasible solutions from DCOPF-based dispatches. The pipeline uses four DCOPF variants and applies AC feasibility recovery using both distributed slack allocation and PV/PQ switching. The main objective is to identify the most effective pipeline for restoring AC feasibility. Evaluation across over \(\mathrm{10{,}000}\) dispatch scenarios on various test cases demonstrates that the structured ACPF model yields solutions that satisfy both the ACPF equations, and all engineering inequality constraints. In a $\mathbf{13{,}659}$-bus case, the mean absolute error and cost differences between $\text{DCOPF}$ and $\text{ACOPF}$ are reduced by 75\% and 93\%, respectively, compared to conventional single slack bus methods. Under extreme loading conditions, the pipeline reduces inequality constraint violations by a factor of 3 to 5.
\end{abstract}

\begin{IEEEkeywords}
AC Feasibility, DC Power Flow (DCPF), Distributed Slack, Loss-Augmented DCOPF, Newton Method.
\end{IEEEkeywords}

\vspace{-0.6em}

\section*{Nomenclature}

\begin{tabbing}
    ${\mathcal{N}}$ \hspace{1.5cm} \= Set of buses; ${\mathcal{N}} = \{1,\, \dots,\, N\}$ \\
    $\mathcal{E}$ \> Set of lines \{lines run from $i \to j$ or $j \to i$\} \\
    $\mathcal{G},\, \mathcal{L}$ \> Set of generators and loads \{$\mathcal{G},\, \mathcal{L} \subseteq \mathcal{N}$\} \\
    $\mathbf{r},\, \mathbf{b},\, \mathbf{g}$ \> Line resistance, susceptance, and conductance  \\ 
    $\mathbf{Y}$ \> Complex branch line admittance \\ 
    $\mathbf{\Phi}$ \> Line-Bus PTDF matrix, size $\mathcal{E} \times \mathcal{N}$\\
    $\mathbf{s}_\mathrm{tx}^{\max}$ \> Apparent power transformer limit \\ 
    $\mathbf{i}_{\mathrm{line}}^{\max}$ \> Current flow line limit \\ 
    $\mathbf{p}_{\mathrm{g}}^{\min},\, \mathbf{p}_{\mathrm{g}}^{\max}$ \> Active power generation limits \\ 
    $\mathbf{p}_{{\mathrm{g}}},\, \mathbf{q}_{{\mathrm{g}}}$ \> Active and reactive power generation \\ 
    $\overrightarrow{{p}_{ij}},\,\overleftarrow{{p}_{ij}}$ \> Active power flows $i\!\to\! j$ and $j\!\to\! i$ \\
    $\mathbf{s}_{\mathrm{g}}$ \> Complex power generation; $\mathbf{p}_{\mathrm{g}} + \mathbf{j} \cdot \mathbf{q}_{\mathrm{g}}$ \\ 
    $\mathbf{p}_{{\mathrm{d}}},\, \mathbf{q}_{{\mathrm{d}}}$ \> Active and reactive power demand \\ 
    $\mathbf{q}_{\mathrm{g}}^{\min},\, \mathbf{q}_{\mathrm{g}}^{\max}$ \> Reactive power generation limits \\ 
    $\mathbf{v}^{\min},\, \mathbf{v}^{\max}$ \> Voltage magnitude limits, for ${v}_i$ \\
    $\boldsymbol{\theta}^{\mathrm{dc}},\, \boldsymbol{\theta}^{\mathrm{ac}}$ \> Voltage angles from DC and AC models \\ 
    $\boldsymbol{\pi}_{\mathrm{g}}$ \> Slack participation factor for generators  \\ 
    $\boldsymbol{\ell}^{\mathrm{tot}},\, \boldsymbol{\epsilon}$ \> Total losses across network, tolerance \\
\end{tabbing}
\vspace{-2em}

\begingroup
  \hypersetup{hidelinks} 
  \thanksto{\noindent\footnotesize The work was partially funded by Los Alamos National Laboratory's Directed Research and Development project, ``Artificial Intelligence for Mission (ArtIMis)'' under U.S. DOE Contract No. DE-AC52-06NA25396. This work was partially funded by NSF award 2112533.\\ 
  }
\endgroup

\section{Introduction} \label{sec:intro}

The linearity of the DC power flow equations provides the DC Optimal Power Flow (DCOPF) problem with significant computational advantages over the AC Optimal Power Flow (ACOPF) problem~\cite{Kile2014_ACvsDC_PSCC}. These advantages come at the cost of accuracy relative to ACOPF solutions. Since inaccuracies in DCOPF solutions can lead to suboptimal operation and violations of operational limits, studying DC power flow accuracy is a long-running research topic that includes both empirical and analytical assessments. For instance,~\cite{Kile2014_ACvsDC_PSCC, liu2002,overbye2004,purchala2005,li2007,stott2009dc,qi2012, TaheriMolzahn_PSCC2024, coffrin2012dc_accuracy,cetinay2017} empirically study DC power flow accuracy for various applications. 

Analytical analyses, e.g.,~\cite{kaye1984, Bolognani-IEEE_Trans, DvijothamMolzahn2016DCBounds}, rigorously bound the worst-case DCPF approximation error, but these bounds are not necessarily indicative of typical accuracy. 

Of particular relevance to this paper, Baker in~\cite{baker2021solutions} analytically proved that, under nonzero loading conditions in networks with positive resistances and reactances, the feasible regions of DCOPF and ACOPF are disjoint. This implies that the optima of conventional DCOPF problems are never AC feasible, i.e., never satisfy the AC power flow equations. The primary argument in~\cite{baker2021solutions} relates to the difference in losses between the DC and AC power flow equations. Baker further shows that even DCOPF variants which incorporate loss-adjustment mechanisms will still fail to provide AC feasible solutions due to a similar argument at the level of individual buses.

The theory in~\cite{baker2021solutions} provides an important foundational result regarding the accuracy of DC power flow approximations. However, this theory does not address how frequently the solutions to DCOPF problems can be \emph{easily restored} to acceptable AC feasible operating points, i.e., operating points which satisfy both the AC power flow equations and the limits on  generator outputs, voltage magnitudes, and line flows. This paper empirically addresses this question.
Solving the AC power flow equations is a natural approach for restoring an AC feasible operating point from a solution to a DCOPF problem. By fixing the generators' active power setpoints to those from the DCOPF solution and the voltage magnitudes to a nominal value such as $1$ per unit, one can easily construct a system of AC power flow equations that are frequently solvable via standard Newton-Raphson methods. The resulting power flow solution is feasible with respect to the AC power flow equations by construction. If the power flow solution satisfies limits on generator outputs, voltage magnitudes,  and line flow limits, it is a feasible operating point with respect to the ACOPF problem's constraints. Thus, while the DCOPF problem's solution is not itself AC feasible as shown in~\cite{baker2021solutions}, it may be the case that an AC feasible solution can often be easily obtained.

Indeed, much of the prior literature on DC power flow accuracy (e.g.,~\cite{Kile2014_ACvsDC_PSCC, liu2002,overbye2004,purchala2005,li2007,stott2009dc,qi2012, TaheriMolzahn_PSCC2024, coffrin2012dc_accuracy,cetinay2017}) and restoration of AC feasible solutions (e.g.,~\cite{zamzam2020learning, zhao2023end, agarwal2018continuous, fang2022ac, TaheriMolzahn2024ACRestoration}) takes this AC power flow approach. However, while existing work performs extensive numerical studies on this topic, there are key gaps in the existing literature. Namely, prior research either \emph{$\mathrm{1}$)}~focuses on simplistic ACPF models that use a single slack bus or neglect the generators' reactive power limits \emph{and/or} \emph{$\mathrm{2}$)}~focuses on DC power flow accuracy over a range of power injections as opposed to the solutions to DCOPF problems. Since DCOPF solutions are usually extreme points in the operating region, approximation accuracy at these points is not necessarily aligned with DC power flow accuracy over the entire operating region. To the best of our knowledge, the only other paper that 
simultaneously considers distributed slack and reactive power limited formulations for ACPF settings similar to ours is~\cite{agarwal2018continuous}. The paper examined reactive power control in ACPF, and addressed a slack distribution formulation -- though independently, and only applies distributed slack on a $\mathrm{23}$-bus system. Further, it does not apply any of the formulations in the context of feasibility restoration for DCOPF solutions.

This work addresses these gaps using a ``structured'' ACPF model that more accurately represent generator behavior via distributed slack bus and PV/PQ switching models by imposing realistic generator limits on both active and reactive power outputs. Using the structured ACPF model with several DCPF approximation variants, results from the $\text{DCOPF}\!\rightarrow\!\text{ACPF}$ pipeline—solving a DCOPF and evaluating the resulting setpoints with an ACPF—demonstrate that different DCPF formulations influence DCOPF dispatch outcomes and the ability to recover AC feasibility with respect to the ACOPF constraints. This provides a crucial empirical counterpart to the theoretical findings of~\cite{baker2021solutions} on the \emph{ease of restoration} for DCOPF solutions. Importantly, the pipeline is a feasibility-recovery procedure and is not intended to emulate an \text{ACOPF} solver. Rigorous recovery bounds are possible in principle but are challenging for the same nonconvex and nonsmooth reasons as ACOPF, so tight relaxations would be costly and simpler ones conservative. Thus, making our empirical results practical and essential.
In summary, the paper’s contributions are:
\begin{itemize}
\item A $\text{DCOPF}\!\rightarrow\!\text{ACPF}$ recovery pipeline identifying the most effective DC formulation for obtaining AC feasible solutions across a variety of large-scale power systems.
\item An assessment of the impacts of ACPF variants—featuring distributed slack and/or PV/PQ switching models—in the context of AC-feasbility restoration.
\item A systematic sensitivity analysis of the pipeline to load perturbations and guidance on its importance for improving feasibility‐restoration workflows.
\end{itemize}
The paper is organized as follows. Section~\ref{sec:pro} introduces the pipeline with \textit{(A)} loss-augmented DCOPF models, and \textit{(B)} a structured ACPF formulation. Section~\ref{sec:num} shows the numerical results, and Section~\ref{sec:sect} concludes with future directions.

\begin{figure}[t] 
\centering

\begin{minipage}{\linewidth}
\noindent\rule{\linewidth}{0.6pt}

\vspace{-0.2em}
\noindent \text{\textbf{Model 1:} DC Optimal Power Flow (DCOPF) Formulation}

\vspace{-0.6em}
\noindent\rule{\linewidth}{0.3pt}

\vspace{-1.3em}

\begin{small}
\begin{subequations}
\label{eq:DCOPF}
\begin{align}
   & \hspace{-0.8em}\min \sum_{i \in \mathcal{G}} \big( c_{2_i}\,p_{g_i}^2 + c_{1_i}\,p_{g_i} + c_{0_i} \big)
&& \text{(Cost Minimization)} \label{eq:DCOPF-cost} \\[0em]
    & \hspace{-0.8em} \textbf{s.t. } p_{\mathrm{g}_i} - p_{\mathrm{d}_i} = \sum_{j \in \mathcal{E}} \overrightarrow{{p}_{ji}} - \sum_{j \in \mathcal{E}} \overrightarrow{{p}_{ij}}
    && \forall i \in \mathcal{N} \text{ (Power Balance)} \label{eq:DCOPF-balance} \\[0em]
    & \overrightarrow{{p}_{ij}} = b_{ij}(\theta_i^{\mathrm{dc}} - \theta_j^{\mathrm{dc}}) 
    && \forall (i,j) \in \mathcal{E} \text{ (DC Flow)} \label{eq:DCOPF-flow} \\
    & |\overrightarrow{{p}_{ij}}| \leq p_{ij}^{\max} 
    && \forall (i,j) \in \mathcal{E} \text{ (Line Limit)} \label{eq:DCOPF-plim} \\
    & \mathbf{p}_{\mathrm{g}}^{\min} \leq \mathbf{p}_{\mathrm{g}} \leq \mathbf{p}_{\mathrm{g}}^{\max}  
    && \text{ (Gen. Active Power)} \label{eq:DCOPF-gen}
\end{align}
\end{subequations}
\end{small}

\vspace{-1em}
\noindent\rule{\linewidth}{0.3pt}
\end{minipage}

\label{fig:opf_models}
 \vspace{-1.1em}
\end{figure}

\begin{figure}[!t]
    \centering
   \includegraphics[width=1\linewidth]{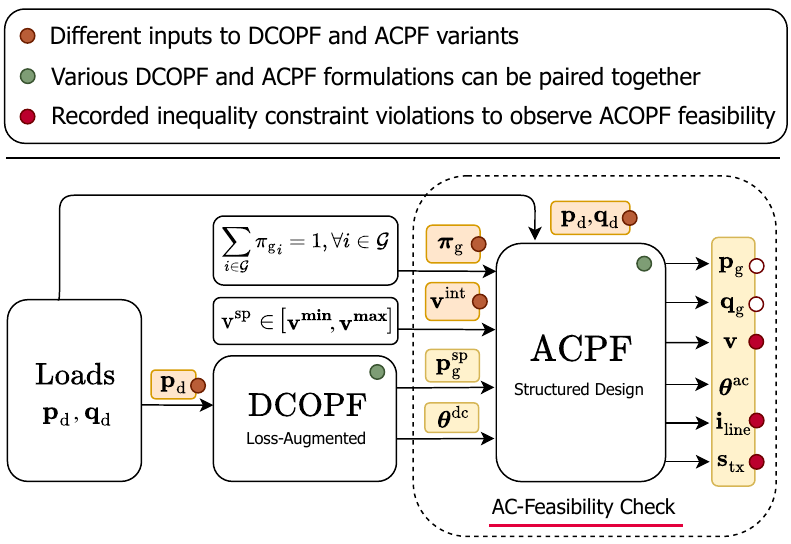}

\vspace{0.2em}

\caption{Pipeline of the $\text{DCOPF}\!\rightarrow\!\text{ACPF}$ model. The DCOPF gives $\mathbf{{p}}_{\mathrm{g}}^\mathrm{sp}$ and $\boldsymbol{\theta}^{\mathrm{dc}}$ from input $\mathbf{p}_{\mathrm{d}}$. The ACPF performs feasibility checks with distributed slack using participation factors ($\boldsymbol{\pi}_{\mathrm{g}}$). Voltage initialization ($\mathbf{v}_{\mathrm{}}^{\mathrm{int}}$) aids in convergence. ACOPF constraints are marked with red.} 

    \label{fig:Pipeline}
     \vspace{-1.2em}
\end{figure}

\vspace{0.5em}

\section{Methodology} \label{sec:pro}

To investigate the most effective $\text{DCOPF}\rightarrow\text{ACPF}$ pipeline as shown in Fig.~\ref{fig:Pipeline}, two components were integrated: (i) a \emph{loss-aware} DCOPF embedding a
loss approximation while remaining convex and efficient, and (ii) a \emph{structured} ACPF with distributed slack and PV/PQ switching. The stages are coupled: the DCOPF provides generator setpoints, and the ACPF applies minimal adjustments to obtain solutions that are nearly or fully AC feasible --- i.e., solutions that satisfy both the ACPF equations and the full ACOPF inequality constraints.


\subsection{DCOPF with Line Loss Approximation} \label{sec:lin}

To improve the accuracy of the DCOPF while preserving tractability, several loss-augmented DCOPF variants have been developed in the literature, three of which are described next.

\textit{1) Linear Line Loss Factor Model (LLLF)} \cite{zhao2023end, coffrin2012approximating}:
Using the DC PTDF ($\boldsymbol{\Phi}$) matrix from the DC power flow approximation \cite{wood1996power}, resistive losses are linearized about a reference DC operating point. Total losses are expressed as:
\vspace{-0.2em}
\begin{equation}
    \boldsymbol{\ell}^{\mathrm{tot}} \;=\; \ell^{\mathrm{ref}} + \boldsymbol{\lambda}^{\top}\mathbf{p},
    \label{eq:lllf_main}
\end{equation}
\vspace{-0.2em}
where $\ell^{\mathrm{ref}}$ is a scalar offset, $\boldsymbol{\lambda}$ is the (linear) loss-factor vector, and $\mathbf{p}$ stacks bus active-power injections. The loss factors follow from differentiating the approximate quadratic loss $\sum_{(i,j)} r_{ij}\,\overrightarrow{p}_{ij}^{2}$ via the chain rule, yielding
$\boldsymbol{\lambda}^{\top} = -2\,(\mathbf{R}\!\odot\!\overrightarrow{\mathbf{p}}^{\mathrm{ref}})^{\top}\boldsymbol{\Phi}$,
and the offset is chosen to match the true loss at the reference,
$\ell^{\mathrm{ref}} = -\boldsymbol{\lambda}^{\top}\mathbf{p}^{\mathrm{ref}} + \sum_{(i,j)\in\mathcal{E}} r_{ij}\,(\overrightarrow{p}_{ij}^{\mathrm{ref}})^{2}$.
Here, $\mathbf{R}$ stacks line resistances $r_{ij}$, $\overrightarrow{\mathbf{p}}^{\mathrm{ref}}$ are oriented reference branch flows, and $\odot$ denotes the Hadamard product \cite{wood1996power}.

\textit{2) Line Loss Quadratic Convex Program (LQCP)} \cite{zhao2023end}:  
This model incorporates line losses directly through convex quadratic constraints. Power balance and operational limits follow~\eqref{eq:DCOPF-plim}–\eqref{eq:DCOPF-gen}. Flows are modeled as being linear in voltage angles via~\eqref{eq:DCOPF-flow}. For each line $(i,j)$, losses are modeled as:
\vspace{-0.2em}
\begin{equation}
\overrightarrow{p_{ij}} + \overleftarrow{p_{ij}} \geq r_{ij} \left( \overrightarrow{p_{ij}} \right)^2.
\label{eq:LQCP_main}
\end{equation}

While this formulation is convex, the presence of nonlinear constraints may increase computational burden~\cite{zhao2023end}.

\textit{3) Line Loss Outer Approximation (LLOA)} \cite{zhao2023end}:  
This approach approximates the nonlinear LQCP loss constraints using supporting hyperplanes. For a given reference flow $\overrightarrow{{p}_{ij}}^{\mathrm{ref}}$, the quadratic loss term is outer-approximated linearly via:
\begin{equation}
    \overrightarrow{{p}_{ij}} + \overleftarrow{{p}_{ij}} \geq r_{ij} \left[ -\left( \overrightarrow{{p}_{ij}}^{\mathrm{ref}} \right)^2 + 2 \overrightarrow{{p}_{ij}}^{\mathrm{ref}} \cdot \overrightarrow{{p}_{ij}} \right],
    \label{eq:lloa_main}
\end{equation}
which makes the approximation conservative with respect to the LQCP formulation. 
The total losses $\sum_{(i,j)} \ell_{ij}$ are embedded into the system power balance via $\sum_i p_i = \sum_d p_d + \sum_{(i,j)} \ell_{ij}$, with each $\ell_{ij}$ lower-bounded by its linear approximation. The LLOA approach allows fast, warm-started dispatch~\cite{zhao2023end}.
\vspace{0.1em}

\subsection{Structured AC Power Flow Model}\label{sec:pri}
Following the DCOPF, a structured ACPF is solved via the Newton–Raphson method, with a distributed slack to allocate active-power imbalances across generators and PV/PQ switching to handle reactive-power limits. For a comprehensive pipeline analysis, a total  of four AC variants are utilized.

\textit{1) Distributed Slack Model:}
A conventional ACPF assigns all active-power imbalance to a single slack generator, often producing unrealistic dispatch and voltage bias. To address this, a \emph{headroom-based} distributed slack formulation is used, sharing the total mismatch $\ell^{\mathrm{tot}}$—the signed active-power deficit or surplus between the DCOPF setpoints ${p_{g_i}^{\mathrm{sp}}}$, and the current AC state $(\boldsymbol{\theta},\mathbf{V})$—across generators in proportion to their available headroom (i.e., margin to their upper active power output limit). This adaptive allocation prevents infeasible loading during AC feasibility recovery, unlike equal or maximum-capacity participation schemes that disregard operating limits. As shown in Model~\ref{eq:HSD}, \eqref{eq:HSD-hi} defines each generator’s headroom $h_i$ as the upward margin to the active-power limit; generators at their upper limits have $h_i=0$ and thus do not absorb further deficit. The participation factor $\pi_{g_i}$ defined in~\eqref{eq:HSD-pi} allocates a fraction of $\ell^{\mathrm{tot}}$ to each unit, and a capacity-proportional fallback ensures feasibility when $\sum_i h_i=0$. Updated injections follow~\eqref{eq:HSD-pg}, with ${\pi_{g_i}}$ recomputed and normalized at each Newton–Raphson iteration to maintain $\pi_{g_i}\ge0$ and $\sum_i\pi_{g_i}=1$. This weighting preserves realism under load perturbations, enhances numerical stability, and improves cost allocation accuracy~\cite{Bharatwaj2012IterativeDCOPF}.

\begin{figure}[!t]
  \vspace{-0.9em}
    \centering
    \includegraphics[width=1\linewidth]{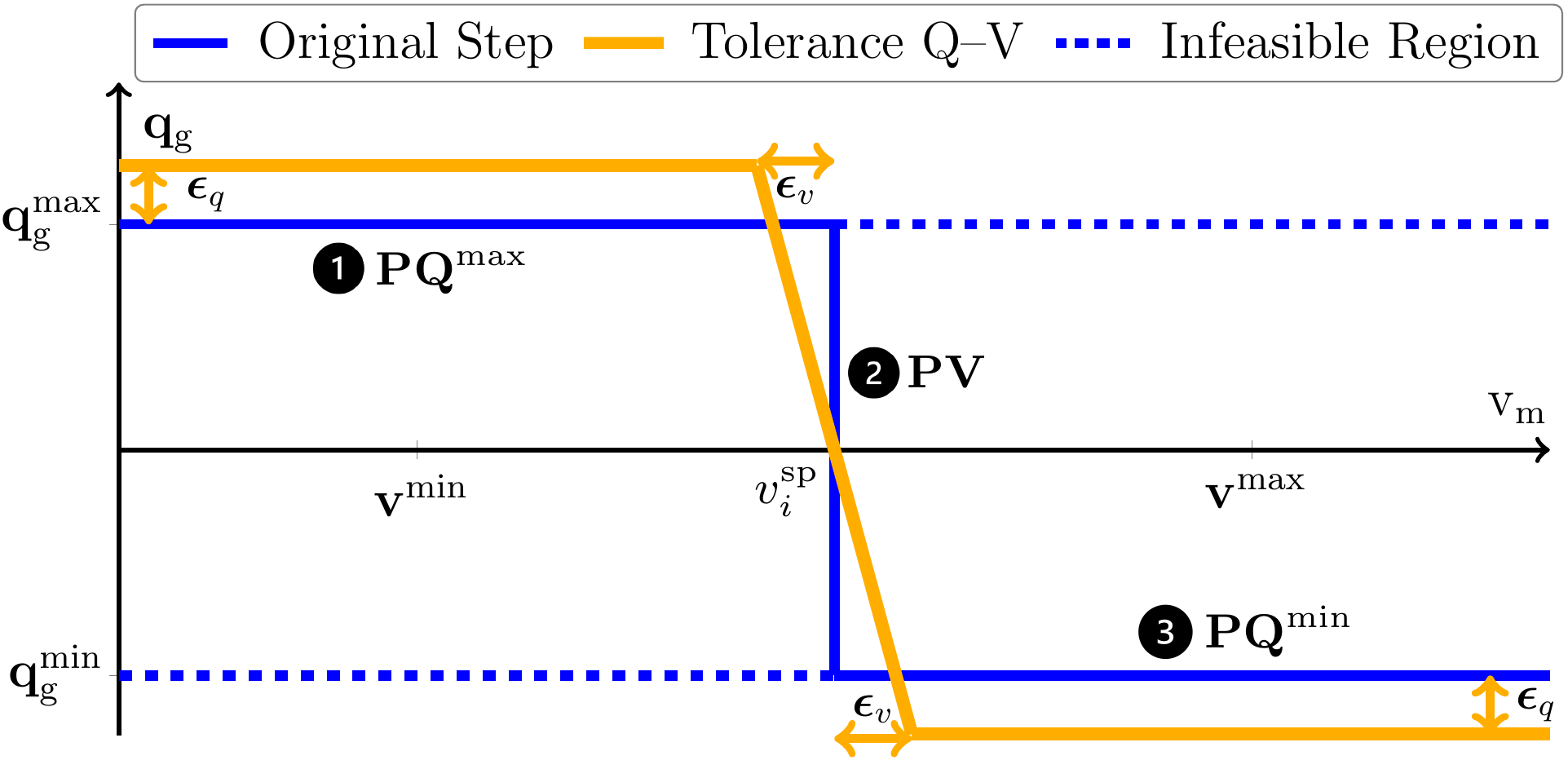}
\vspace{-1em}

\caption{The generator bus-type switching logic. The \textcolor{blue}{\textbf{blue curve}} is the discrete logic. The bus transitions between $\mathbf{PQ}^{\mathrm{max}}$, $\mathbf{PV}$, and $\mathbf{PQ}^{\mathrm{min}}$ states, based on ${q_g}_i$ and $v_i^{\mathrm{sp}}$. The \textcolor{blue}{\textbf{blue dashed lines}} are infeasible regions where the generator cannot maintain voltage control while obeying reactive power limits. The \textcolor{orange}{\textbf{orange curve}} is the tolerance-based control, with tolerances $\mathrm{\epsilon}_q$ and $\mathrm{\epsilon}_v$ on reactive power and voltage.}

    
   \label{fig:bts}
   \vspace{-1.2em}
\end{figure}

\label{sec:qv}
\textit{2) Reactive Power Control:} The ACPF also utilizes a reactive power control mechanism via bus-type switching---a standard method for representing generator behavior in power flow analysis. As shown in Fig.~\ref{fig:bts}, the bus transitions between $\mathbf{PQ}^{\max}$, $\mathbf{PV}$, and $\mathbf{PQ}^{\min}$ states based on $(q_{g_i}, v_i^{\mathrm{sp}})$. In practice, most solvers apply small tolerances to avoid oscillatory switching during Newton–Raphson iterations. Accordingly, a deadband $(\epsilon_q, \epsilon_v)$ is imposed such that $|\mathbf{q}_{\mathbf{g}}-\mathbf{q}_{\mathbf{g}}^{\mathrm{lim}}|\le\epsilon_q$ and $|\mathbf{v}^*-\mathbf{v}^{\mathrm{sp}}|\le\epsilon_v$, preventing transitions within this range. When reactive power violations exceed $\epsilon_q$, $\mathbf{PV}$ buses switch to $\mathbf{PQ}$, clamping reactive power at its bound. Reversion occurs only when voltages move beyond $\epsilon_v$. This practical scheme limits switching per Newton solve, enforces reactive power feasibility, and improves convergence by reducing oscillations. Although it does not guarantee voltage-limit satisfaction, it maintains feasible reactive injections and numerically stable operation. 

Unstable and erratic iterations from discrete PV/PQ switching noted in prior studies~\cite{zhao2008bus,zeng2023pvpq} were rarely observed in $\mathrm{AC}_{\text{SPF}}$, and while homotopy-based methods~\cite{agarwal2018continuous} can further reduce iterations, the tolerance-based deadband offers sufficient robustness and integrates seamlessly with the ACPF framework.

In Newton-Raphson ACPF, \emph{total iteration count} records all inner iterations needed for convergence under fixed bus types, while the \emph{PV/PQ switching count} tracks outer iterations where generator buses change type as reactive limits are enforced or released. Both iteration counts are recorded in Section \ref{sec:num}.

For the sake of notational brevity, the paper hereafter denote four ACPF variants as follows: $\mathrm{AC}_{\text{BASE}}$ uses a single slack bus with no generator control, 
$\mathrm{AC}_{\text{BTS}}$ applies discrete PV/PQ switching (blue curve in Fig.~\ref{fig:bts}), 
$\mathrm{AC}_{\text{DS}}$ employs distributed slack control only, and 
$\mathrm{AC}_{\text{SPF}}$ (\textbf{S}tructured \textbf{P}ower \textbf{F}low) combines headroom-based distributed slack (Model~\ref{eq:HSD}) with tolerance-based PV/PQ switching (orange curve in Fig.~\ref{fig:bts}).

\begin{figure}[!t]
\noindent\rule{\linewidth}{0.6pt}
\vspace{-0.4em}
\noindent\text{\textbf{Model 2:} Headroom-Based Slack Distribution}
\vspace{-0.5em}
\noindent\rule{\linewidth}{0.3pt}
\vspace{-1.2em}
\begin{small} 
\setcounter{equation}{1}
\refstepcounter{equation}
\label{eq:HSD}
\setcounter{equation}{4}
\begin{subequations}
\begin{align}
h_i &= \max\!\big(p_{g_i}^{\max} - p_{g_i}^{\mathrm{sp}},\,0\big),
&& \!\!\!\!\forall i \in \mathcal{G}
&& \hspace{-1em}\text{\,(Headroom)}
\label{eq:HSD-hi}\\[-0.1em]
\pi_{g_i} &=
\begin{cases}
   \displaystyle \frac{h_i}{\sum_{i \in \mathcal{G}} h_i},  \text{~if } \sum_{i \in \mathcal{G}} h_i > 0,\\[6pt]
   \displaystyle \frac{p_{g_i}^{\max}}{\sum_{i \in \mathcal{G}} p_{g_i}^{\max}},  \text{~otherwise,}
\end{cases}
&& \!\!\!\!\forall i \in \mathcal{G}
&& \hspace{-1em}\text{\,(Slack share)}
\label{eq:HSD-pi}\\[-0.1em]
p_{g_i} &= p_{g_i}^{\mathrm{sp}} + \pi_{g_i}\,\ell^{\mathrm{tot}},
&& \!\!\!\!\forall i \in \mathcal{G}
&& \hspace{-1em}\text{\,(New dispatch)}
\label{eq:HSD-pg}
\end{align}
\end{subequations}
\end{small}

\vspace{-0.8em}
\noindent\rule{\linewidth}{0.3pt}
\end{figure}

\begin{table*}[!htpb]
\centering
\renewcommand{\arraystretch}{0.2}
\caption{{\small\scshape Number of Violations in Base Case for DCOPF $\rightarrow$ ACPF Pipelines Using Standard Test Cases}}
\scriptsize
\setlength{\tabcolsep}{2pt}
\resizebox{\linewidth}{!}{
\begin{tabular}{llcccccccccccccccc}
\toprule
\textbf{Test case} & \textbf{AC type}
& \multicolumn{4}{c}{\textbf{Active Power Violations}}
& \multicolumn{4}{c}{\textbf{Reactive Power Violations}}
& \multicolumn{4}{c}{\textbf{Voltage Violations}}
& \multicolumn{4}{c}{\textbf{Thermal Violations}} \\
\cmidrule(lr){3-6} \cmidrule(lr){7-10} \cmidrule(lr){11-14} \cmidrule(lr){15-18}
& 
& $\scriptsize\mathrm{DC}_{\text{BASE}}$
& $\scriptsize\mathrm{DC}_{\text{LLLF}}$
& $\scriptsize\mathrm{DC}_{\text{LQCP}}$
& $\scriptsize\mathrm{DC}_{\text{LLOA}}$
& $\scriptsize\mathrm{DC}_{\text{BASE}}$
& $\scriptsize\mathrm{DC}_{\text{LLLF}}$
& $\scriptsize\mathrm{DC}_{\text{LQCP}}$
& $\scriptsize\mathrm{DC}_{\text{LLOA}}$
& $\scriptsize\mathrm{DC}_{\text{BASE}}$
& $\scriptsize\mathrm{DC}_{\text{LLLF}}$
& $\scriptsize\mathrm{DC}_{\text{LQCP}}$
& $\scriptsize\mathrm{DC}_{\text{LLOA}}$
& $\scriptsize\mathrm{DC}_{\text{BASE}}$
& $\scriptsize\mathrm{DC}_{\text{LLLF}}$
& $\scriptsize\mathrm{DC}_{\text{LQCP}}$
& $\scriptsize\mathrm{DC}_{\text{LLOA}}$ \\
\midrule
\midrule
\multirow{4}{*}{\texttt{case\_118}} & $\mathrm{AC}_{\text{BASE}}$
& $\mathrm{1}$ & $\mathrm{1}$ & $\mathrm{0}$ & $\mathrm{1}$
& $\mathrm{22}$ & $\mathrm{21}$ & $\mathrm{18}$ & $\mathrm{16}$
& $\mathrm{0}$ & $\mathrm{1}$ & $\mathrm{0}$ & $\mathrm{0}$
& $\mathrm{0}$ & $\mathrm{0}$ & $\mathrm{0}$ & $\mathrm{0}$ \\
& $\mathrm{AC}_{\text{BTS}}$
& $\mathrm{1}$ & $\mathrm{1}$ & $\mathrm{0}$ & $\mathrm{1}$
& $\mathrm{0}$ & $\mathrm{0}$ & $\mathrm{0}$ & $\mathrm{0}$
& $\mathrm{1}$ & $\mathrm{1}$ & $\mathrm{0}$ & $\mathrm{0}$
& $\mathrm{1}$ & $\mathrm{1}$ & $\mathrm{1}$ & $\mathrm{1}$ \\
& $\mathrm{AC}_{\text{DS}}$
& $\mathrm{0}$ & $\mathrm{0}$ & $\mathrm{0}$ & $\mathrm{0}$
& $\mathrm{19}$ & $\mathrm{20}$ & $\mathrm{25}$ & $\mathrm{26}$
& $\mathrm{0}$ & $\mathrm{1}$ & $\mathrm{0}$ & $\mathrm{0}$
& $\mathrm{0}$ & $\mathrm{0}$ & $\mathrm{0}$ & $\mathrm{0}$ \\
& $\mathrm{AC}_{\text{SPF}}$
& $\mathbf{0}$ & $\mathbf{0}$ & $\mathbf{0}$ & $\mathbf{0}$
& $\mathbf{0}$ & $\mathbf{0}$ & $\mathbf{0}$ & $\mathbf{0}$
& $\mathbf{0}$ & $\mathbf{1}$ & $\mathbf{0}$ & $\mathbf{0}$
& $\mathbf{0}$ & $\mathbf{0}$ & $\mathbf{0}$ & $\mathbf{0}$ \\
\midrule
\multirow{4}{*}{\texttt{case\_1354}} & $\mathrm{AC}_{\text{BASE}}$
& $\mathrm{1}$ & $\mathrm{1}$ & $\mathrm{1}$ & $\mathrm{1}$
& $\mathrm{91}$ & $\mathrm{70}$ & $\mathrm{85}$ & $\mathrm{76}$
& $\mathrm{2}$ & $\mathrm{1}$ & $\mathrm{1}$ & $\mathrm{1}$
& $\mathrm{3}$ & $\mathrm{17}$ & $\mathrm{5}$ & $\mathrm{11}$ \\
& $\mathrm{AC}_{\text{BTS}}$
& $\mathrm{1}$ & $\mathrm{1}$ & $\mathrm{0}$ & $\mathrm{1}$
& $\mathrm{0}$ & $\mathrm{0}$ & $\mathrm{0}$ & $\mathrm{0}$
& $\mathrm{1}$ & $\mathrm{1}$ & $\mathrm{1}$ & $\mathrm{1}$
& $\mathrm{6}$ & $\mathrm{13}$ & $\mathrm{5}$ & $\mathrm{10}$ \\
& $\mathrm{AC}_{\text{DS}}$
& $\mathrm{0}$ & $\mathrm{0}$ & $\mathrm{0}$ & $\mathrm{0}$
& $\mathrm{82}$ & $\mathrm{67}$ & $\mathrm{87}$ & $\mathrm{89}$
& $\mathrm{1}$ & $\mathrm{1}$ & $\mathrm{1}$ & $\mathrm{0}$
& $\mathrm{5}$ & $\mathrm{15}$ & $\mathrm{1}$ & $\mathrm{4}$ \\
& $\mathrm{AC}_{\text{SPF}}$
& $\mathbf{0}$ & $\mathbf{0}$ & $\mathbf{0}$ & $\mathbf{0}$
& $\mathbf{0}$ & $\mathbf{0}$ & $\mathbf{0}$ & $\mathbf{0}$
& $\mathbf{0}$ & $\mathbf{1}$ & $\mathbf{1}$ & $\mathbf{0}$
& $\mathbf{0}$ & $\mathbf{13}$ & $\mathbf{2}$ & $\mathbf{0}$ \\
\midrule
\multirow{4}{*}{\texttt{case\_2869}} & $\mathrm{AC}_{\text{BASE}}$
& $\mathrm{1}$ & $\mathrm{1}$ & $\mathrm{1}$ & $\mathrm{1}$
& $\mathrm{310}$ & $\mathrm{190}$ & $\mathrm{87}$ & $\mathrm{115}$
& $\mathrm{2}$ & $\mathrm{1}$ & $\mathrm{1}$ & $\mathrm{1}$
& $\mathrm{12}$ & $\mathbf{4}$ & $\mathrm{3}$ & $\mathrm{16}$ \\
& $\mathrm{AC}_{\text{BTS}}$
& $\mathrm{1}$ & $\mathrm{1}$ & $\mathrm{1}$ & $\mathrm{1}$
& $\mathrm{0}$ & $\mathrm{0}$ & $\mathrm{0}$ & $\mathrm{0}$
& $\mathrm{14}$ & $\mathrm{7}$ & $\mathrm{4}$ & $\mathrm{2}$
& $\mathrm{12}$ & $\mathrm{25}$ & $\mathrm{13}$ & $\mathrm{22}$ \\
& $\mathrm{AC}_{\text{DS}}$
& $\mathrm{0}$ & $\mathrm{0}$ & $\mathrm{0}$ & $\mathrm{0}$
& $\mathrm{365}$ & $\mathrm{213}$ & $\mathrm{164}$ & $\mathrm{96}$
& $\mathrm{2}$ & $\mathrm{2}$ & $\mathrm{3}$ & $\mathrm{2}$
& $\mathrm{18}$ & $\mathrm{17}$ & $\mathrm{5}$ & $\mathrm{11}$ \\
& $\mathrm{AC}_{\text{SPF}}$
& $\mathbf{0}$ & $\mathbf{0}$ & $\mathbf{0}$ & $\mathbf{0}$
& $\mathbf{0}$ & $\mathbf{0}$ & $\mathbf{0}$ & $\mathbf{0}$
& $\mathbf{0}$ & $\mathbf{1}$ & $\mathbf{0}$ & $\mathbf{0}$
& $\mathbf{0}$ & $\mathrm{6}$ & $\mathbf{1}$ & $\mathbf{0}$ \\
\midrule
\multirow{4}{*}{\texttt{case\_13659}} & $\mathrm{AC}_{\text{BASE}}$
& $\mathrm{1}$ & $\mathrm{1}$ & $\mathrm{1}$ & $\mathrm{0}$
& $\mathrm{1215}$ & $\mathrm{1028}$ & $\mathrm{971}$ & $\mathrm{1012}$
& $\mathrm{19}$ & $\mathrm{6}$ & $\mathrm{6}$ & $\mathrm{5}$
& $\mathrm{68}$ & $\mathrm{65}$ & $\mathrm{62}$ & $\mathrm{62}$ \\
& $\mathrm{AC}_{\text{BTS}}$
& $\mathrm{1}$ & $\mathrm{1}$ & $\mathrm{0}$ & $\mathrm{1}$
& $\mathrm{0}$ & $\mathrm{0}$ & $\mathrm{0}$ & $\mathrm{0}$
& $\mathrm{213}$ & $\mathrm{170}$ & $\mathrm{69}$ & $\mathrm{62}$
& $\mathrm{93}$ & $\mathrm{55}$ & $\mathrm{52}$ & $\mathrm{60}$ \\
& $\mathrm{AC}_{\text{DS}}$
& $\mathrm{0}$ & $\mathrm{0}$ & $\mathrm{0}$ & $\mathrm{0}$
& $\mathrm{1003}$ & $\mathrm{980}$ & $\mathrm{956}$ & $\mathrm{940}$
& $\mathrm{22}$ & $\mathrm{16}$ & $\mathrm{17}$ & $\mathrm{16}$
& $\mathrm{40}$ & $\mathrm{42}$ & $\mathrm{32}$ & $\mathrm{33}$ \\
& $\mathrm{AC}_{\text{SPF}}$
& $\mathbf{0}$ & $\mathbf{0}$ & $\mathbf{0}$ & $\mathbf{0}$
& $\mathbf{0}$ & $\mathbf{0}$ & $\mathbf{0}$ & $\mathbf{0}$
& $\mathbf{0}$ & $\mathbf{7}$ & $\mathbf{0}$ & $\mathbf{3}$
& $\mathbf{0}$ & $\mathbf{24}$ & $\mathbf{0}$ & $\mathbf{21}$ \\
\bottomrule
\end{tabular}}
\label{tab:violation-count}
\vspace{-0.4em}
\end{table*}

\begin{table*}[!htpb]
\centering
\renewcommand{\arraystretch}{0.4}
\caption{{\small\scshape Maximum Violation Value in Base Case for DCOPF $\rightarrow$ ACPF Pipelines Using Standard Test Cases}}
\scriptsize
\setlength{\tabcolsep}{3.4pt}
\resizebox{\linewidth}{!}{
\begin{tabular}{l l cccc cccc cccc cccc}
\toprule
\textbf{Test case} & \textbf{AC type}
& \multicolumn{4}{c}{\textbf{Active Power Violations (p.u.)}}
& \multicolumn{4}{c}{\textbf{Reactive Power Violations (p.u.)}}
& \multicolumn{4}{c}{\textbf{Voltage Violations (p.u.)}}
& \multicolumn{4}{c}{\textbf{Thermal Violations (\%)}} \\
\cmidrule(lr){3-6} \cmidrule(lr){7-10} \cmidrule(lr){11-14} \cmidrule(lr){15-18}
& & $\scriptsize \mathrm{DC}_{\text{BASE}}$ & $\scriptsize \mathrm{DC}_{\text{LLLF}}$ & $\scriptsize \mathrm{DC}_{\text{LQCP}}$ & $\scriptsize \mathrm{DC}_{\text{LLOA}}$
  & $\scriptsize \mathrm{DC}_{\text{BASE}}$ & $\scriptsize \mathrm{DC}_{\text{LLLF}}$ & $\scriptsize \mathrm{DC}_{\text{LQCP}}$ & $\scriptsize \mathrm{DC}_{\text{LLOA}}$
  & $\scriptsize \mathrm{DC}_{\text{BASE}}$ & $\scriptsize \mathrm{DC}_{\text{LLLF}}$ & $\scriptsize \mathrm{DC}_{\text{LQCP}}$ & $\scriptsize \mathrm{DC}_{\text{LLOA}}$
  & $\scriptsize \mathrm{DC}_{\text{BASE}}$ & $\scriptsize \mathrm{DC}_{\text{LLLF}}$ & $\scriptsize \mathrm{DC}_{\text{LQCP}}$ & $\scriptsize \mathrm{DC}_{\text{LLOA}}$ \\
\midrule
\midrule
\multirow{4}{*}{\texttt{case\_118}}
& $\mathrm{AC}_{\text{BASE}}$
& $\mathrm{0.002}$ & $\mathrm{0.001}$ & $\mathrm{0.000}$ & $\mathrm{0.001}$
& $\mathrm{1.538}$ & $\mathrm{1.528}$ & $\mathrm{1.489}$ & $\mathrm{1.469}$
& $\mathrm{0.000}$ & $\mathrm{0.010}$ & $\mathrm{0.000}$ & $\mathrm{0.000}$
& $\mathrm{0.000}$ & $\mathrm{0.000}$ & $\mathrm{0.000}$ & $\mathrm{0.000}$ \\
& $\mathrm{AC}_{\text{BTS}}$
& $\mathrm{0.001}$ & $\mathrm{0.001}$ & $\mathrm{0.000}$ & $\mathrm{0.001}$
& $\mathrm{0.000}$ & $\mathrm{0.000}$ & $\mathrm{0.000}$ & $\mathrm{0.000}$
& $\mathrm{0.008}$ & $\mathrm{0.005}$ & $\mathrm{0.000}$ & $\mathrm{0.000}$
& $\mathrm{4.472}$ & $\mathrm{4.470}$ & $\mathrm{4.100}$ & $\mathrm{4.001}$ \\
& $\mathrm{AC}_{\text{DS}}$
& $\mathrm{0.000}$ & $\mathrm{0.000}$ & $\mathrm{0.000}$ & $\mathrm{0.000}$
& $\mathrm{1.234}$ & $\mathrm{1.202}$ & $\mathrm{1.113}$ & $\mathrm{1.110}$
& $\mathrm{0.000}$ & $\mathrm{0.020}$ & $\mathrm{0.000}$ & $\mathrm{0.000}$
& $\mathrm{0.000}$ & $\mathrm{0.000}$ & $\mathrm{0.000}$ & $\mathrm{0.000}$ \\
& $\mathrm{AC}_{\text{SPF}}$
& $\mathbf{0.000}$ & $\mathbf{0.000}$ & $\mathbf{0.000}$ & $\mathbf{0.000}$
& $\mathbf{0.000}$ & $\mathbf{0.000}$ & $\mathbf{0.000}$ & $\mathbf{0.000}$
& $\mathbf{0.000}$ & $\mathbf{0.005}$ & $\mathbf{0.000}$ & $\mathbf{0.000}$
& $\mathbf{0.000}$ & $\mathbf{0.000}$ & $\mathbf{0.000}$ & $\mathbf{0.000}$ \\
\midrule
\multirow{4}{*}{\texttt{case\_1354}}
& $\mathrm{AC}_{\text{BASE}}$
& $\mathrm{0.024}$ & $\mathrm{0.016}$ & $\mathrm{0.012}$ & $\mathrm{0.012}$
& $\mathrm{14.29}$ & $\mathrm{15.78}$ & $\mathrm{13.72}$ & $\mathrm{13.88}$
& $\mathrm{0.030}$ & $\mathrm{0.024}$ & $\mathrm{0.016}$ & $\mathrm{0.009}$
& $\mathrm{14.16}$ & $\mathrm{12.59}$ & $\mathrm{11.73}$ & $\mathrm{11.83}$ \\
& $\mathrm{AC}_{\text{BTS}}$
& $\mathrm{0.014}$ & $\mathrm{0.013}$ & $\mathrm{0.000}$ & $\mathrm{0.014}$
& $\mathrm{0.000}$ & $\mathrm{0.000}$ & $\mathrm{0.000}$ & $\mathrm{0.000}$
& $\mathrm{0.017}$ & $\mathbf{0.015}$ & $\mathrm{0.014}$ & $\mathrm{0.012}$
& $\mathrm{18.85}$ & $\mathrm{16.60}$ & $\mathrm{17.14}$ & $\mathrm{19.07}$ \\
& $\mathrm{AC}_{\text{DS}}$
& $\mathrm{0.000}$ & $\mathrm{0.000}$ & $\mathrm{0.000}$ & $\mathrm{0.000}$
& $\mathrm{12.78}$ & $\mathrm{12.35}$ & $\mathrm{12.10}$ & $\mathrm{12.12}$
& $\mathrm{0.021}$ & $\mathrm{0.016}$ & $\mathrm{0.011}$ & $\mathrm{0.000}$
& $\mathrm{15.42}$ & $\mathbf{11.23}$ & $\mathrm{8.433}$ & $\mathrm{9.761}$ \\
& $\mathrm{AC}_{\text{SPF}}$
& $\mathbf{0.000}$ & $\mathbf{0.000}$ & $\mathbf{0.000}$ & $\mathbf{0.000}$
& $\mathbf{0.000}$ & $\mathbf{0.000}$ & $\mathbf{0.000}$ & $\mathbf{0.000}$
& $\mathbf{0.000}$ & $\mathrm{0.020}$ & $\mathbf{0.010}$ & $\mathbf{0.000}$
& $\mathbf{0.000}$ & $\mathrm{15.65}$ & $\mathbf{3.810}$ & $\mathbf{0.000}$ \\
\midrule
\multirow{4}{*}{\texttt{case\_2869}}
& $\mathrm{AC}_{\text{BASE}}$
& $\mathrm{2.522}$ & $\mathrm{1.714}$ & $\mathrm{1.131}$ & $\mathrm{1.362}$
& $\mathrm{14.51}$ & $\mathrm{14.62}$ & $\mathrm{14.08}$ & $\mathrm{14.53}$
& $\mathrm{0.026}$ & $\mathbf{0.020}$ & $\mathrm{0.021}$ & $\mathrm{0.024}$
& $\mathrm{27.56}$ & $\mathrm{17.42}$ & $\mathrm{12.53}$ & $\mathrm{11.67}$ \\
& $\mathrm{AC}_{\text{BTS}}$
& $\mathrm{2.121}$ & $\mathrm{1.032}$ & $\mathrm{2.116}$ & $\mathrm{1.089}$
& $\mathrm{0.000}$ & $\mathrm{0.000}$ & $\mathrm{0.000}$ & $\mathrm{0.000}$
& $\mathrm{0.039}$ & $\mathrm{0.027}$ & $\mathrm{0.023}$ & $\mathrm{0.025}$
& $\mathrm{24.37}$ & $\mathrm{19.62}$ & $\mathrm{14.09}$ & $\mathrm{20.96}$ \\
& $\mathrm{AC}_{\text{DS}}$
& $\mathrm{0.000}$ & $\mathrm{0.000}$ & $\mathrm{0.000}$ & $\mathrm{0.000}$
& $\mathrm{16.02}$ & $\mathrm{14.39}$ & $\mathrm{11.76}$ & $\mathrm{11.76}$
& $\mathrm{0.033}$ & $\mathrm{0.020}$ & $\mathrm{0.017}$ & $\mathrm{0.002}$
& $\mathrm{26.78}$ & $\mathrm{14.34}$ & $\mathrm{9.051}$ & $\mathrm{9.803}$ \\
& $\mathrm{AC}_{\text{SPF}}$
& $\mathbf{0.000}$ & $\mathbf{0.000}$ & $\mathbf{0.000}$ & $\mathbf{0.000}$
& $\mathbf{0.000}$ & $\mathbf{0.000}$ & $\mathbf{0.000}$ & $\mathbf{0.000}$
& $\mathbf{0.000}$ & $\mathrm{0.024}$ & $\mathbf{0.000}$ & $\mathbf{0.000}$
& $\mathbf{0.000}$ & $\mathbf{6.732}$ & $\mathbf{4.329}$ & $\mathbf{0.000}$ \\
\midrule
\multirow{4}{*}{\texttt{case\_13659}}
& $\mathrm{AC}_{\text{BASE}}$
& $\mathrm{43.10}$ & $\mathrm{42.05}$ & $\mathrm{44.30}$ & $\mathrm{0.000}$
& $\mathrm{8.790}$ & $\mathrm{7.322}$ & $\mathrm{6.999}$ & $\mathrm{6.704}$
& $\mathrm{0.063}$ & $\mathrm{0.055}$ & $\mathrm{0.053}$ & $\mathrm{0.042}$
& $\mathrm{55.45}$ & $\mathrm{56.01}$ & $\mathrm{48.67}$ & $\mathrm{46.34}$ \\
& $\mathrm{AC}_{\text{BTS}}$
& $\mathrm{44.52}$ & $\mathrm{43.46}$ & $\mathrm{0.000}$ & $\mathrm{33.80}$
& $\mathrm{0.000}$ & $\mathrm{0.000}$ & $\mathrm{0.000}$ & $\mathrm{0.000}$
& $\mathrm{0.150}$ & $\mathrm{0.129}$ & $\mathrm{0.112}$ & $\mathrm{0.122}$
& $\mathrm{75.43}$ & $\mathrm{70.13}$ & $\mathrm{62.24}$ & $\mathrm{67.60}$ \\
& $\mathrm{AC}_{\text{DS}}$
& $\mathrm{0.000}$ & $\mathrm{0.000}$ & $\mathrm{0.000}$ & $\mathrm{0.000}$
& $\mathrm{9.334}$ & $\mathrm{7.592}$ & $\mathrm{6.178}$ & $\mathrm{6.207}$
& $\mathrm{0.071}$ & $\mathrm{0.067}$ & $\mathrm{0.066}$ & $\mathrm{0.061}$
& $\mathrm{46.73}$ & $\mathrm{41.28}$ & $\mathrm{40.01}$ & $\mathrm{40.13}$ \\
& $\mathrm{AC}_{\text{SPF}}$
& $\mathbf{0.000}$ & $\mathbf{0.000}$ & $\mathbf{0.000}$ & $\mathbf{0.000}$
& $\mathbf{0.000}$ & $\mathbf{0.000}$ & $\mathbf{0.000}$ & $\mathbf{0.000}$
& $\mathbf{0.000}$ & $\mathbf{0.008}$ & $\mathbf{0.000}$ & $\mathbf{0.007}$
& $\mathbf{0.000}$ & $\mathbf{15.65}$ & $\mathbf{0.000}$ & $\mathbf{13.81}$ \\
\bottomrule
\end{tabular}}
\label{tab:max-violations}
\vspace{-0.5em}
\end{table*}

\section{Numerical Experiments} \label{sec:num}
\hspace{0.5em} The various combinations of DCOPFs and ACPFs described in Section~\ref{sec:pro} were evaluated on a diverse suite of M{\sc atpower}~8.1 test cases \cite{zimmerman2011matpower}. This includes $\mathrm{IEEE}$ cases \{\texttt{ieee\_30}, \texttt{ieee\_ne\_39}, \texttt{ieee\_118}\}, $\mathrm{ACTIVSg}$ grids \{\texttt{South\_Carolina\_500}, \texttt{Texas\_2000}\}, $\mathrm{PEGASE}$ networks \{\texttt{pegase\_89}, \texttt{pegase\_1354}, \texttt{pegase\_2869}, \texttt{pegase\_9241}, \texttt{pegase\_13659}\}, and a large-scale $\mathrm{RTE}$ case \{\texttt{rte\_6468}\}. Computations were carried out on the Darwin high-performance computing system at Los Alamos National Laboratory. Experiments used a single $\mathrm{24}$-core compute node equipped with $\mathrm{32}$ GB of RAM. The ACOPFs and DCOPFs variants were solved using \texttt{PowerModels.jl}~\cite{coffrin2018powermodels}. The design and implementation of the custom ACPF was formulated using \texttt{pandapower.py}~\cite{thurner2018pandapower}. Active power demands ($\mathbf{p}_\mathrm{d}$) at load buses were perturbed using Gaussian multiplicative noise, where each positive demand was scaled by $\boldsymbol{\xi} \sim \mathcal{N}(1.0, \boldsymbol{\sigma}^2)$ with $\boldsymbol{\sigma}\in({5\%,15\%})$. Reactive power ($\mathbf{q}_\mathrm{d}$) was recomputed from randomly sampled power factors in $[0.95,1.0]$, and $1000$ samples were evaluated per testcase. Since distributed slack and bus-type switching can change the Jacobian's structure across iterations, we rebuild the sparse system when the active set changes while exploiting sparsity at each step. These optimizations yield a scalable Newton-Raphson solver. The solver tolerance was set to $10^{-6}$ p.u., while the tolerances for reactive power control were $10^{-4}$ p.u. for $\mathrm{\epsilon}_q$ (reactive power) and $10^{-5}$ p.u. for $\mathrm{\epsilon}_v$ (voltage). These values were chosen from prior runs and empirical evaluations; future work will study systematic tuning and sensitivity analysis. All results reported correspond to samples that converged (average convergence rate $>80\%$). The mean absolute error (MAE) and percent cost difference (CD) are computed as follows:
\vspace{-0.2em}
\begin{equation} \label{eq:MAE}
    \mathbf{MAE} = \frac{1}{{G}} 
    \left\lVert \mathbf{p}_\mathrm{g}^{\, \mathrm{DC} \rightarrow \mathrm{AC}} 
    - \mathbf{p}_\mathrm{g}^{\, \mathcal{O}} \right\rVert_{1},
\end{equation}
\vspace{-0.5em}
\begin{equation}\label{eq:cost_error}
\mathbf{CD}
= \frac{\left| \mathrm{Cost}^{\,\mathrm{DC}\!\rightarrow\!\mathrm{AC}} 
- \mathrm{Cost}^{\mathcal{O}} \right|}
       {\mathrm{Cost}^{\mathcal{O}}}\cdot 100,
\end{equation}

\noindent where $G$ is the number of generators, $\mathbf{p}_\mathrm{g}^{\, \mathrm{DC} \rightarrow \mathrm{AC}}$ is the generation vector after the $\text{DCOPF}\!\rightarrow\!\text{ACPF}$ pipeline, and $\mathbf{p}_\mathrm{g}^{\, \mathcal{O}}$ is the ACOPF-based reference used as a ground truth.

\subsection{ACOPF Feasibility Study with $\textit{DCOPF}\!\rightarrow\!\textit{ACPF}$ Variants} \label{sec:ACP}

This section evaluates violations of the ACOPF constraints in Fig.~\ref{fig:Pipeline} across the different $\text{DCOPF}\rightarrow\text{ACPF}$ pipelines. Tables and charts summarize these violations and their reduction factors under loss-augmented DCOPFs and the ACPF variants.

\textit{1) Violation Tables Analysis:}
Tables~\ref{tab:violation-count} and \ref{tab:max-violations} show that the distributed–slack reconciliation $\mathrm{AC}_{\text{SPF}}$ delivers the fewest violations across all DC initializations and systems. In \texttt{case\_13659}, reactive-power violations are noticeably high for $\mathrm{AC}_{\text{BASE}}$ and $\mathrm{AC}_{\text{DS}}$, since these lack reactive control. The \emph{counts} are $1215/1028/971/1012$ for $\mathrm{AC}_{\text{BASE}}$ across $\mathrm{DC}_{\text{BASE}}/\mathrm{DC}_{\text{LLLF}}/\mathrm{DC}_{\text{LQCP}}/\mathrm{DC}_{\text{LLOA}}$, and $1003/980/956/940$ for $\mathrm{AC}_{\text{DS}}$. By contrast, both $\mathrm{AC}_{\text{BTS}}$ and $\mathrm{AC}_{\text{SPF}}$ keep reactive violations at zero, with $\mathrm{AC}_{\text{SPF}}$ also maintaining voltage violations low ($0/7/0/3$) and thermal violations modest ($0/24/0/21$). For active-power violations, the distributed slack in $\mathrm{AC}_{\text{DS}}$ and $\mathrm{AC}_{\text{SPF}}$ again drives violations to zero across all DC variants. However, $\mathrm{AC}_{\text{BASE}}$ and $\mathrm{AC}_{\text{BTS}}$ struggle: in \texttt{case\_13659}, maximum per-unit active-power violations reach $43.10/42.05/44.30/0.000$\,p.u.\ under $\mathrm{AC}_{\text{BASE}}$ and $44.52/43.46/0.000/33.80$\,p.u.\ under $\mathrm{AC}_{\text{BTS}}$. For $\mathrm{AC}_{\text{SPF}}$, maximum active- and reactive-power violations remain $0.000$\,p.u.\ in every case and DC variant. Thus, $\mathrm{AC}_{\text{SPF}}$ combines the best of both worlds, eliminating active and reactive violations. $\mathrm{AC}_{\text{SPF}}$ also seems to restore feasibility for $\mathrm{DC}_{\text{BASE}}$ consistently. This will be revisited in the sensitivity analysis, later in this section.

Looking horizontally within any fixed AC reconciliation, violations generally fall as we move from the lossless $\mathrm{DC}_{\text{BASE}}$ to the lossy $\mathrm{DC}_{\text{LQCP}}/\mathrm{DC}_{\text{LLOA}}$. $\mathrm{DC}_{\text{LLLF}}$ usually improves on $\mathrm{DC}_{\text{BASE}}$ but less reliably. As summarized in Section~\ref{sec:pro}, the superior accuracy of $\mathrm{DC}_{\text{LQCP}}$ and $\mathrm{DC}_{\text{LLOA}}$ arises because they include lossy formulations that more faithfully capture $\mathrm{I^2R}$ effects, whereas $\mathrm{DC}_{\text{LLLF}}$, being a simpler linear formulation, neglects higher-order terms and therefore underperforms. Ultimately, when observing the $\mathrm{AC}_{\text{SPF}}$ results horizontally for each case, the $\mathrm{DC}_{\text{LQCP}}$ was revealed to generalize better and be more consistent with violation reduction, at scale. $\mathrm{DC}_{\text{LLOA}}$ did have the lowest violations sparingly compared to the others, but performed poorly at scale (e.g., for \texttt{case\_13659} in $\mathrm{AC}_{\text{SPF}}$, complete AC feasibility was achieved in $\mathrm{DC}_{\text{LQCP}}$, with $\mathrm{DC}_{\text{LLOA}}$ recording violations closer to the linearized $\mathrm{DC}_{\text{LLLF}}$. 

\begin{figure*}[!t]
    \centering
    \includegraphics[width=1\textwidth]{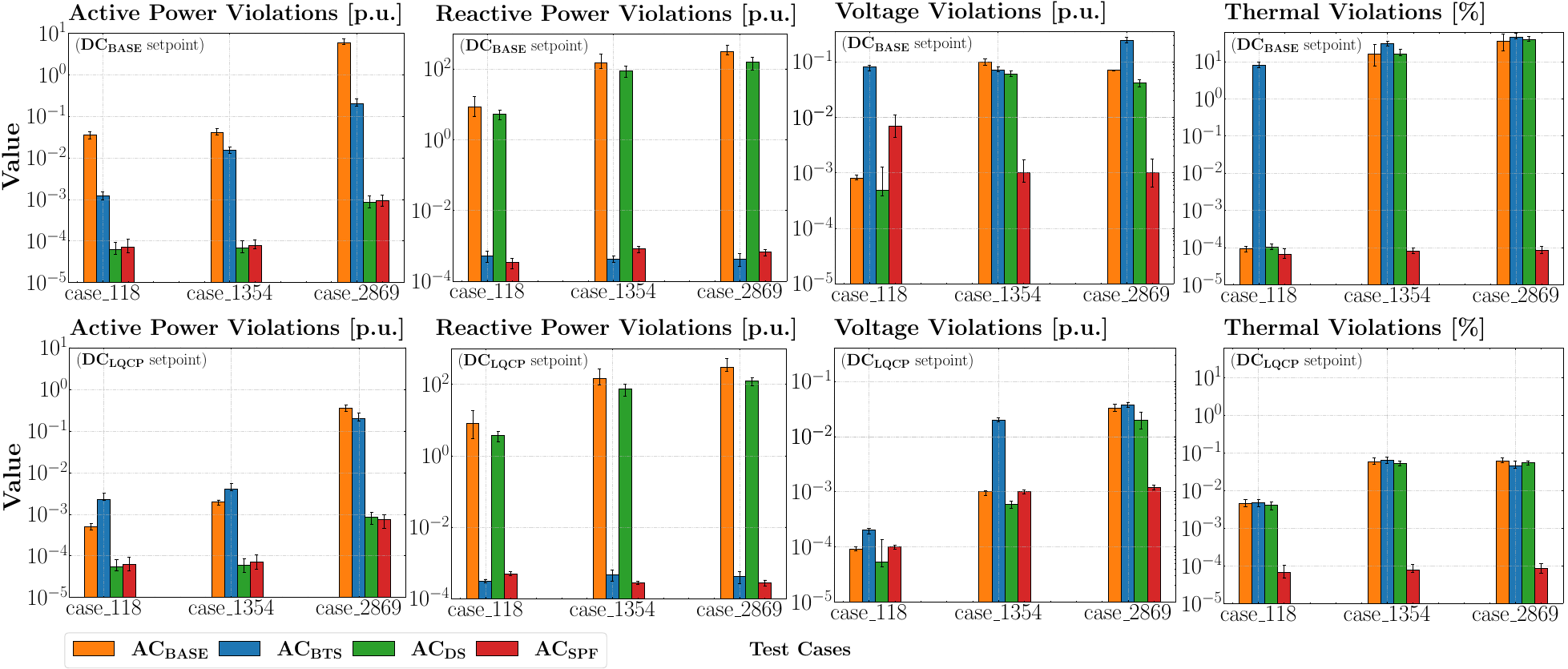}
\caption{
Comparison of the sum total of violations in different AC power flow formulations across various cases: \texttt{case\_118}, \texttt{case\_1354}, and \texttt{case\_2869} (with mean and min–max error bars). The pipelines shown have different DC setpoints: (top) {\scriptsize $\mathrm{DC}_{\text{BASE}}$} and
(bottom) {\scriptsize $\mathrm{DC}_{\text{LQCP}}$}. The plots are taken over load uncertainty with ${\sigma}=15\%$, for $\mathrm{1000}$ samples per case. 
Shown are the active power, reactive power, voltage, and thermal violations, with y-axes on log-scales.
Results are reported for {\scriptsize $\mathrm{AC}_{\text{BASE}}$} (\textbf{\textcolor{orange}{orange}}),
{\scriptsize $\mathrm{AC}_{\text{BTS}}$} (\textbf{\textcolor{blue}{blue}}), {\scriptsize $\mathrm{AC}_{\text{DS}}$} (\textbf{\textcolor{ACgreen}{green}}), and
{\scriptsize $\mathrm{AC}_{\text{SPF}}$} (\textbf{\textcolor{red}{red}}). 
}
\label{fig:violations_dcbase_dcl2}
\vspace{-0.5em}
\end{figure*}

\textit{2) Bar Graph Sensitivity Analysis:} 
\label{sec:sensitivity}
Following the realization of $\mathrm{DC}_{\text{LQCP}}$'s better generalizeable performance, imperative sensitivity analysis was conducted. Fig.~\ref{fig:violations_dcbase_dcl2} provides a visual reflection of the numerical trends, presenting the \emph{sum of violation magnitudes} across all buses, branches, and limits. Unlike Tables~\ref{tab:violation-count}–\ref{tab:max-violations}, which details violation \emph{counts} and \emph{maxima}, these plots capture the aggregate severity. The top panels correspond to $\mathrm{DC}_{\text{BASE}} \!\rightarrow\! \mathrm{AC}$ pipelines, while the bottom panels use lossy $\mathrm{DC}_{\text{LQCP}}$ initializations. Across most test systems, the distributed–slack $\mathrm{AC}_{\text{SPF}}$ reconciliation yields the lowest summed violations in every category. From the upper plots, moving from $\mathrm{AC}_{\text{BASE}}$ to $\mathrm{AC}_{\text{SPF}}$ reduces both active- and reactive-power violation magnitudes by roughly \emph{four orders of magnitude} (from $10^{-3}$--$10^{2}$, \ down to $10^{-4}$--$10^{-3}$ p.u.). The same trend extends to voltage and thermal violations, which remain low under $\mathrm{AC}_{\text{SPF}}$. In contrast, $\mathrm{AC}_{\text{BTS}}$ and $\mathrm{AC}_{\text{BASE}}$ show higher aggregate magnitudes. $\mathrm{AC}_{\text{DS}}$ performed well in active power and voltage violation reduction, but performed poorly in reducing reactive power and thermal violations.

Comparing the upper and lower rows, lossy initializations ($\mathrm{DC}_{\text{LQCP}}$) further reduce total violation magnitudes by an additional one-to-two orders of magnitude across all AC types. The $\mathrm{DC}_{\text{LQCP}}$ formulation captures loss and angle effects more accurately than $\mathrm{DC}_{\text{BASE}}$, producing better-aligned initializations and greater AC feasibility. Consequently, the $\mathrm{DC}_{\text{LQCP}}\! \rightarrow\! \mathrm{AC}_{\text{SPF}}$ pipeline is the most consistent at violation reduction across test cases. These plots reinforce the trends in Tables~\ref{tab:violation-count}–\ref{tab:max-violations}: incorporating lossy DC models and distributed slack improves both constraint satisfaction and numerical stability.

\begin{table*}[!t]
\centering
\renewcommand{\arraystretch}{1}
\caption{{\small\scshape Performance Metrics in Base Case DCOPF for ACPF Variants Across Multiple Test Cases}}
\scriptsize
\setlength{\tabcolsep}{3pt}
\resizebox{\linewidth}{!}{
\begin{tabular}{
l
S[table-format=1.2] S[table-format=1.2] S[table-format=1.2] S[table-format=1.2] c
S[table-format=1.2] S[table-format=1.2] S[table-format=1.2] S[table-format=1.2] c
c c c c
c c
}
\toprule
\textbf{Test case}
& \multicolumn{5}{c}{\textbf{Cost Difference (\%)}} 
& \multicolumn{5}{c}{\textbf{Mean Absolute Error (p.u.)}} 
& \multicolumn{4}{c}{\textbf{Iteration Count (Total)}} 
& \multicolumn{2}{c}{\textbf{Solving Time (s)}}\\
\cmidrule(lr){2-6} \cmidrule(lr){7-11} \cmidrule(lr){12-15} \cmidrule(lr){16-17}
& {\scriptsize $\mathrm{AC}_{\text{BASE}}$}
& {\scriptsize $\mathrm{AC}_{\text{BTS}}$}
& {\scriptsize $\mathrm{AC}_{\text{DS}}$}
& {\scriptsize $\mathrm{AC}_{\text{SPF}}$}
& {\scriptsize \textbf{Improv. (\%)}}
& {\scriptsize $\mathrm{AC}_{\text{BASE}}$}
& {\scriptsize $\mathrm{AC}_{\text{BTS}}$}
& {\scriptsize $\mathrm{AC}_{\text{DS}}$}
& {\scriptsize $\mathrm{AC}_{\text{SPF}}$}
& {\scriptsize \textbf{Improv. (\%)}}
& {\scriptsize $\mathrm{AC}_{\text{BASE}}$}
& {\scriptsize $\mathrm{AC}_{\text{BTS}}^{\star}$}
& {\scriptsize $\mathrm{AC}_{\text{DS}}$}
& {\scriptsize $\mathrm{AC}_{\text{SPF}}^{\star}$}
& {\scriptsize ACOPF}
& {\scriptsize $\mathrm{AC}_{\text{SPF}}$} \\
\midrule
\midrule
\texttt{case\_30}    
& $\mathrm{1.21}$ & $\mathrm{1.71}$ & $\mathrm{0.36}$ & $\mathrm{0.32}$ & $(\mathbf{74},\,\mathbf{81},\,\mathbf{11})$
& $\mathrm{0.00}$ & $\mathrm{0.02}$ & $\mathrm{0.01}$ & $\mathrm{0.00}$ & $(\mathrm{0},\,\mathbf{100},\,\mathbf{100})$
& $\mathrm{4}$ & $\mathrm{19}$ & $\mathrm{4}$ & $\mathbf{\mathrm{3}}$
& $\mathrm{1.79}$ & $\mathrm{0.21}$ \\

\texttt{case\_39}    
& $\mathrm{1.38}$ & $\mathrm{1.43}$ & $\mathrm{0.30}$ & $\mathrm{0.26}$ & $(\mathbf{81},\,\mathbf{82},\,\mathbf{13})$
& $\mathrm{0.01}$ & $\mathrm{0.03}$ & $\mathrm{0.01}$ & $\mathrm{0.00}$ & $(\mathbf{100},\,\mathbf{100},\,\mathbf{100})$
& $\mathrm{3}$ & $\mathrm{20}$ & $\mathrm{3}$ & $\mathbf{\mathrm{8}}$
& $\mathrm{0.41}$ & $\mathrm{0.13}$ \\

\texttt{case\_89}    
& $\mathrm{0.33}$ & $\mathrm{0.36}$ & $\mathrm{0.19}$ & $\mathrm{0.20}$ & $(\mathbf{39},\,\mathbf{44},\,-5)$
& $\mathrm{0.02}$ & $\mathrm{0.04}$ & $\mathrm{0.02}$ & $\mathrm{0.00}$ & $(\mathbf{100},\,\mathbf{100},\,\mathbf{100})$
& $\mathrm{4}$ & $\mathrm{42}$ & $\mathrm{4}$ & $\mathbf{\mathrm{12}}$
& $\mathrm{0.42}$ & $\mathrm{0.23}$ \\

\midrule
\texttt{case\_118}   
& $\mathrm{2.19}$ & $\mathrm{2.82}$ & $\mathrm{0.24}$ & $\mathrm{0.18}$ & $(\mathbf{92},\,\mathbf{94},\,\mathbf{25})$
& $\mathrm{0.01}$ & $\mathrm{0.04}$ & $\mathrm{0.00}$ & $\mathrm{0.01}$ & $(\mathrm{0},\,\mathbf{75},\,-100)$
& $\mathrm{4}$ & $\mathrm{60}$ & $\mathbf{\mathrm{3}}$ & $\mathrm{18}$
& $\mathrm{0.19}$ & $\mathrm{0.22}$ \\

\texttt{case\_1354}  
& $\mathrm{0.41}$ & $\mathrm{0.58}$ & $\mathrm{0.12}$ & $\mathrm{0.28}$ & $(\mathbf{32},\,\mathbf{52},\,-133)$
& $\mathrm{0.01}$ & $\mathrm{0.05}$ & $\mathrm{0.03}$ & $\mathrm{0.02}$ & $(\mathrm{0},\,\mathbf{60},\,\mathbf{33})$
& $\mathrm{4}$ & $\mathrm{357}$ & $\mathbf{\mathrm{3}}$ & $\mathrm{152}$
& $\mathrm{2.78}$ & $\mathrm{64.10}$ \\

\texttt{case\_2000}$^{\diamond}$
& $\mathrm{2.11}$ & $\mathrm{2.14}$ & $\mathrm{0.91}$ & $\mathrm{0.82}$ & $(\mathbf{61},\,\mathbf{62},\,\mathbf{10})$
& $\mathrm{0.23}$ & $\mathrm{0.20}$ & $\mathrm{0.02}$ & $\mathrm{0.05}$ & $(\mathbf{78},\,\mathbf{75},\,-150)$
& $\mathbf{\mathrm{3}}$ & $\mathrm{208}$ & $\mathrm{4}$ & $\mathrm{81}$
& $\mathrm{3.24}$ & $\mathrm{36.10}$ \\ 

\midrule
\texttt{case\_2869}  
& $\mathrm{0.54}$ & $\mathrm{0.62}$ & $\mathrm{0.31}$ & $\mathrm{0.37}$ & $(\mathbf{31},\,\mathbf{40},\,-19)$
& $\mathrm{0.04}$ & $\mathrm{0.14}$ & $\mathrm{0.03}$ & $\mathrm{0.04}$ & $(\mathrm{0},\,\mathbf{71},\,-33)$
& $\mathrm{4}$ & $\mathrm{777}$ & $\mathbf{\mathrm{3}}$ & $\mathrm{380}$
& $\mathrm{4.88}$ & $\mathrm{187}$ \\

\texttt{case\_6468}  
& $\mathrm{1.65}$ & $\mathrm{1.89}$ & $\mathrm{0.26}$ & $\mathrm{0.24}$ & $(\mathbf{85},\,\mathbf{87},\,\mathbf{8})$
& $\mathrm{0.07}$ & $\mathrm{0.12}$ & $\mathrm{0.04}$ & $\mathrm{0.05}$ & $(\mathbf{29},\,\mathbf{58},\,-25)$
& $\mathrm{4}$ & $\mathrm{3396}$ & $\mathrm{4}$ & $\mathbf{\mathrm{1670}}$
& $\mathrm{59.05}$ & $\mathrm{2213}$ \\

\texttt{case\_9241}  
& $\mathrm{1.60}$ & $\mathrm{1.11}$ & $\mathrm{0.21}$ & $\mathrm{0.20}$ & $(\mathbf{88},\,\mathbf{82},\,\mathbf{5})$
& $\mathrm{0.02}$ & $\mathrm{0.13}$ & $\mathrm{0.02}$ & $\mathrm{0.02}$ & $(\mathrm{0},\,\mathbf{85},\,\mathrm{0})$
& $\mathbf{\mathrm{3}}$ & $\mathrm{2204}$ & $\mathbf{\mathrm{3}}$ & $\mathrm{1356}$
& $\mathrm{30.18}$ & $\mathrm{6104}$ \\

\texttt{case\_13659} 
& $\mathrm{1.81}$ & $\mathrm{1.31}$ & $\mathrm{0.15}$ & $\mathrm{0.13}$ & $(\mathbf{93},\,\mathbf{90},\,\mathbf{13})$
& $\mathrm{0.01}$ & $\mathrm{0.04}$ & $\mathrm{0.01}$ & $\mathrm{0.01}$ & $(\mathrm{0},\,\mathbf{75},\,\mathrm{0})$
& $\mathrm{4}$ & $\mathrm{6452}$ & $\mathbf{\mathrm{3}}$ & $\mathrm{3320}$
& $\mathrm{127.14}$ & $\mathrm{7170}$ \\
\bottomrule\multicolumn{17}{l}{$\star$ AC variants have higher iteration counts due to sequential PV/PQ switching. $\diamond$ \texttt{case\_2000} incurred violations of inequality constraints.}
\end{tabular}}
\label{tab:loss-functions2}
\vspace{-1.3em}
\end{table*}

\begin{figure*}[!htpb]
    \centering
    \includegraphics[width=1\textwidth]{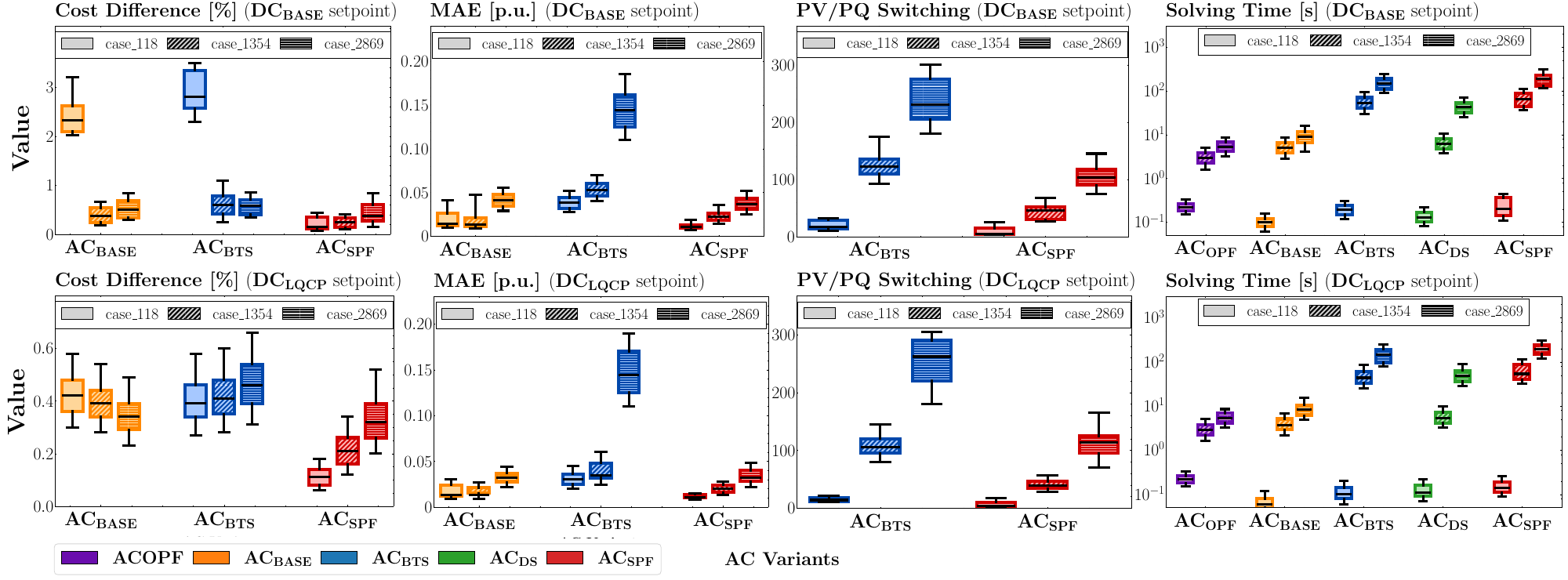}
\caption{Boxplot comparison of different AC variants' metrics across test cases: \texttt{case\_118}, \texttt{case\_1354}, and \texttt{case\_2869} (with mean and min–max extremes). The AC variants shown have different DC setpoints: (top) {\scriptsize $\mathrm{DC}_{\text{BASE}}$} and
(bottom) {\scriptsize $\mathrm{DC}_{\text{LQCP}}$}. The plots are taken over load uncertainty with ${\sigma}=15\% $, for $\mathrm{1000}$ samples per case. Metrics shown (left to right): cost difference, mean absolute error, iteration count, and solving time.
Results are reported for
{\scriptsize $\mathrm{AC}_{\text{OPF}}$} (\textbf{\textcolor{violet}{purple}}),
{\scriptsize $\mathrm{AC}_{\text{BASE}}$} (\textbf{\textcolor{orange}{orange}}),
{\scriptsize $\mathrm{AC}_{\text{BTS}}$} (\textbf{\textcolor{blue}{blue}}),
{\scriptsize $\mathrm{AC}_{\text{DS}}$} (\textbf{\textcolor{ACgreen}{green}}), and
{\scriptsize $\mathrm{AC}_{\text{SPF}}$} (\textbf{\textcolor{red}{red}}).}
\vspace{-0.5em}

    \label{fig:box_plot_comp}
        \vspace{-0.5em}
\end{figure*}

\subsection{Key Metrics for ACPF Variants Assessment} \label{sec:and}

The following study gives a extensive comparative analysis of the ACPF variants from Section~\ref{sec:pro}. After demonstrating that $\mathrm{DC}_{\text{LQCP}}\! \rightarrow\! \mathrm{AC}_{\text{SPF}}$ reduces violations and often fully restores AC feasibility, further performance analysis is required.

\textit{1) Performance Metrics Table:}
Table~\ref{tab:loss-functions2} summarizes key performance metrics for 10 core test cases, comparing cost deviation, mean absolute error, iteration behavior, and solving time across all ACPF variants. Clear trends show that distributed--slack formulations, $\mathrm{AC}_{\text{DS}}$ and $\mathrm{AC}_{\text{SPF}}$, yield the lowest cost deviation and error magnitudes through improved active-power balancing. Cost differences drop by up to $93\%$ relative to $\mathrm{AC}_{\text{BASE}}$, while mean absolute errors approach zero for most systems (e.g., $\mathrm{AC}_{\text{SPF}}$ maintains $\leq 0.05$\,p.u.). Iteration metrics reflect the effects of tolerance-based PV/PQ switching: both $\mathrm{AC}_{\text{BTS}}$ and $\mathrm{AC}_{\text{SPF}}$ reach zero reactive power violations, though $\mathrm{AC}_{\text{SPF}}$ consistently converges faster---reducing total iterations by about $50$\% in \texttt{case\_2869}. The use of sequential bus-type switching\footnote{Sequential PV/PQ switching updates one violated generator per round, whereas grouped switching updates multiple violated generators per round. Grouped switching therefore reduces the number of switching rounds.} in both $\mathrm{AC}_{\text{BTS}}$ and $\mathrm{AC}_{\text{SPF}}$ significantly increased the number of iteration counts and solving times in Table~\ref{tab:loss-functions2}; Table~\ref{tab:grouped_switching_appendix} in the appendix reveals how the use of grouped PV/PQ switching provides improvements to the iteration count and solving time. In contrast, $\mathrm{AC}_{\text{BASE}}$ and $\mathrm{AC}_{\text{DS}}$ complete in a single recalculation loop but require longer inner iterations, at times. Full feasibility was not restored only for \texttt{case\_2000}, likely due to high line impedance ratios $(\mathbf{r}/\mathbf{x})$ limiting reactive controllability. As feasibility restoration is the pipeline's focus, convergence rate was not prioritized, though future work will address it \cite{zeng2023pvpq}. Solving times align with iteration behavior: smaller systems (e.g., \texttt{case\_89}, \texttt{case\_118}) run comparably or slightly faster under $\mathrm{AC}_{\text{SPF}}$, while larger networks slow down due to iterative refinement. Still, the gains in cost and feasibility confirm $\mathrm{AC}_{\text{SPF}}$ has a good balance between accuracy and computational efficiency.


\textit{2) Box Plot Sensitivity Analysis:}
Fig.~\ref{fig:box_plot_comp} illustrates the statistical distribution of performance metrics for three cases, revealing the robustness and variability of each ACPF variant. The upper plots correspond to $\mathrm{DC}_{\text{BASE}}\! \rightarrow\! \mathrm{AC}$ pipelines, while the lower plots represent $\mathrm{DC}_{\text{LQCP}}$ setpoints. The lossy DC generally yields a lower average cost difference and MAE, though with slightly higher variability across realizations. The $\mathrm{DC}_{\text{LQCP}}$ initialization also reduces iteration counts and solving times—particularly for \texttt{case\_118} and \texttt{case\_1354}—owing to its pre-accounting of power losses and generator redispatch. Since the AC reconciliation must only distribute the residual loss (i.e., remaining mismatch after DC loss approximation), convergence is faster. However, reactive-power control adjustments still dominate iteration requirements in larger networks. Across all metrics, $\mathrm{AC}_{\text{SPF}}$ achieves the tightest distributions with the lowest mean values for both cost difference and MAE. The reduced interquartile ranges in $\mathrm{AC}_{\text{SPF}}$ indicate superior consistency and robustness. For iteration count and solving time, $\mathrm{AC}_{\text{SPF}}$ exhibits apparent higher computational effort, but remains competitive—with improved accuracy and reliability.

\begin{figure*}[!t]
  \centering
  \begin{subfigure}{0.49\textwidth}
    \centering
    \includegraphics[width=\linewidth]{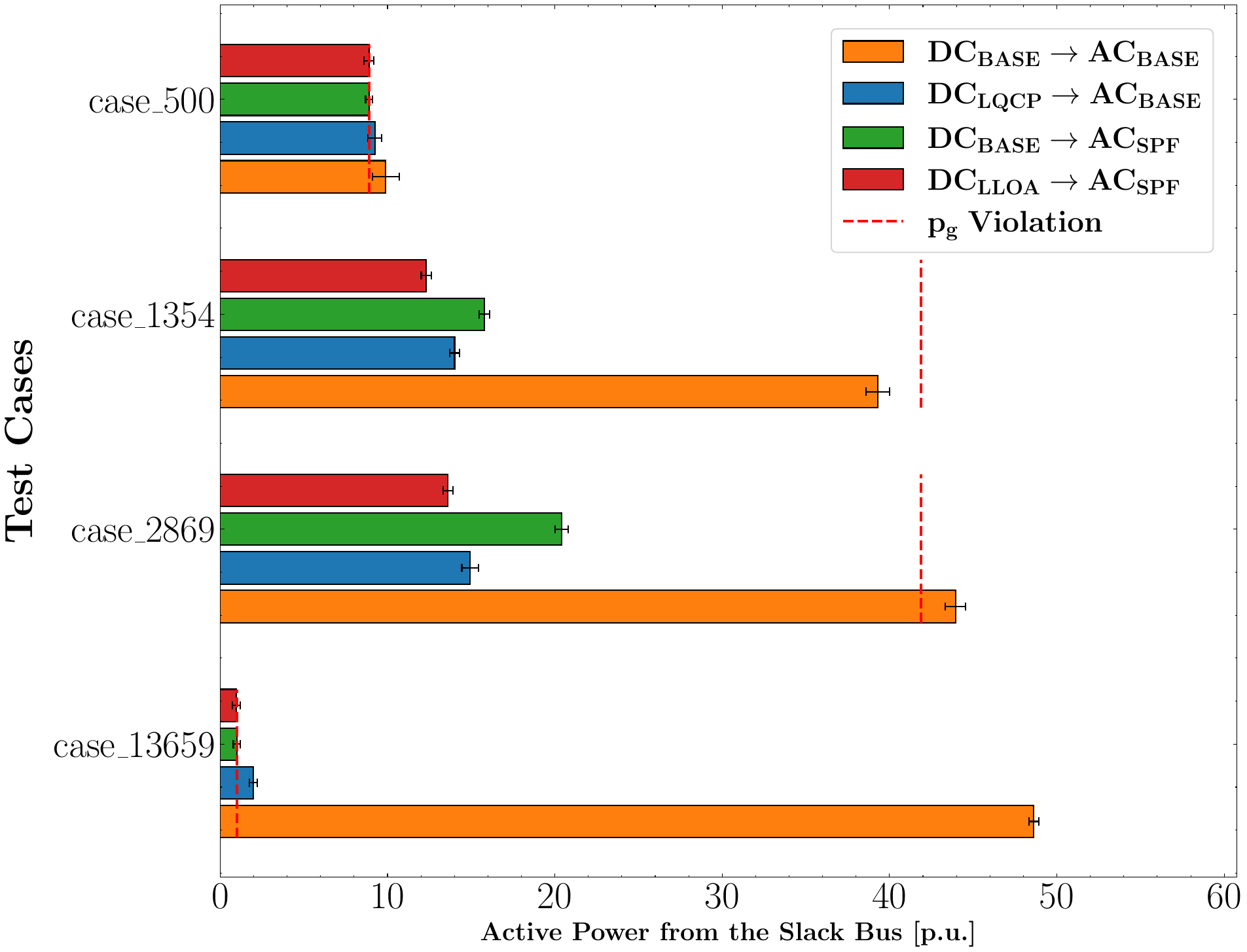}
    \caption{\textbf{Active power from the slack-bus (mean with min–max error bars)}}
    \label{fig:pipeline-a}
  \end{subfigure}\hfill
  \begin{subfigure}{0.49\textwidth}
    \centering
    \includegraphics[width=\linewidth]{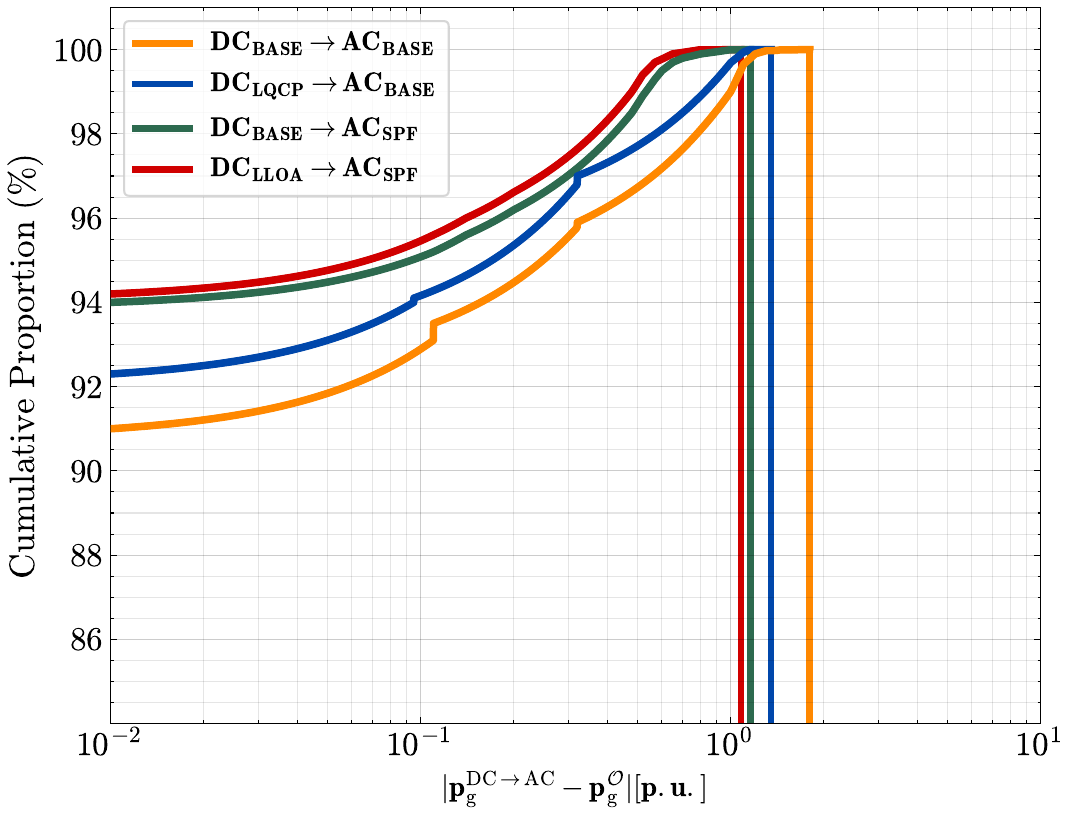}
    \caption{\textbf{Cumulative plot of generator output errors in a \texttt{case\_1354}}}
    \label{fig:pipeline-b}
  \end{subfigure}
  \vspace{-0.2em}
\caption{Pipeline comparison across different DC and AC variants, over load uncertainty with ${\sigma}=5\%$.
    (a) Slack-bus active power using $\mathrm{100}$ per case, for:
    \texttt{case\_500}, \texttt{case\_1354}, \texttt{case\_2869}, and \texttt{case\_13659} under
    {\scriptsize $\mathrm{DC}_{\text{BASE}}\!\rightarrow\!\mathrm{AC}_{\text{BASE}}$}
    (\textbf{\textcolor{orange}{orange}}),
    {\scriptsize $\mathrm{DC}_{\text{LQCP}}\!\rightarrow\!\mathrm{AC}_{\text{BASE}}$}
    (\textbf{\textcolor{blue}{blue}}),
    {\scriptsize $\mathrm{DC}_{\text{BASE}}\!\rightarrow\!\mathrm{AC}_{\text{SPF}}$}
    (\textbf{\textcolor{ACgreen}{green}}), and 
    {\scriptsize $\mathrm{DC}_{\text{LLOA}}\!\rightarrow\!\mathrm{AC}_{\text{SPF}}$}
    (\textbf{\textcolor{red}{red}}). Dashed (\textbf{\textcolor{red}{red}}) line is the active power violation.
    (b) Cumulative density factor (CDF) plot of unit-wise active-power deviation 
    $|\mathbf{p}_\mathrm{g}^{\mathrm{DC}\rightarrow\mathrm{AC}} - \mathbf{p}_\mathrm{g}^{\mathcal{O}}|$ across generators in \texttt{case\_1354}, using $\mathrm{1000}$ per case, with an x-axis log-scale.}
\label{fig:pipeline_comparison}
\vspace{-0.5em}
\end{figure*}

 \subsection{Distributed Slack Effects on Active Power} \label{sec:DCO}

The analysis examines distributed- versus single-slack-bus modeling in ACPF, isolating how DC and AC variants separately affect loss handling. It also evaluates whether DC models with higher degrees of freedom paired with simpler AC models (and vice versa) can yield acceptable performances.

\textit{1) Bar Plot Sensitivity Analysis:}
The horizontal bar plots in Fig.~\ref{fig:pipeline_comparison}(a) reveal significant differences in slack-bus active-power requirements across four pipeline combinations. The red and green pipelines using $\mathrm{AC}_{\text{SPF}}$ consistently yield active-power violations below $10^{-3}\,\text{p.u.}$ Among them, $\mathrm{DC}_\text{LLOA}\rightarrow\mathrm{AC}_{\text{SPF}}$ produces the smallest active-power changes across all test cases. This behavior arises from the distributed-slack mechanism, which shares losses in proportion to generator reserve capability, rather than a single slack bus. By contrast, $\mathrm{DC}_{\text{BASE}}\rightarrow\mathrm{AC}_{\text{BASE}}$ exhibits the largest slack-bus injections, since a single slack must absorb losses after a lossless DCOPF—pushing loss accountability to that unit. The pipeline $\mathrm{DC}_\text{LQCP}\rightarrow\mathrm{AC}_{\text{BASE}}$ also employs a single slack, \emph{but} its setpoints originate from a DCOPF that accounts for losses, yielding a notable reduction relative to $\mathrm{DC}_{\text{BASE}}$. Consequently, using $\mathrm{DC}_\text{LQCP}$ or $\mathrm{DC}_\text{LLOA}$ improves $\mathrm{AC}$ reconciliation compared with $\mathrm{DC}_{\text{BASE}}$ (seen in the reduction from orange to blue, and green to red). With low perturbations of ${\sigma}=5\%$, single-slack pipelines may converge without active-power violations, yet the slack generator remains stressed, underscoring its inherent fragility.

\textit{2) Cumulative Density Function Analysis:}
Fig.~\ref{fig:pipeline_comparison}(b) compares 
$\bigl\lvert \mathbf{p}_{\mathrm{g}}^{\mathrm{DC}\!\rightarrow\!\mathrm{AC}}
 - \mathbf{p}_{\mathrm{g}}^{\mathcal{O}} \bigr\rvert$ (p.u.) 
across pipelines on a logarithmic scale. The two distributed-slack recoveries,
$\mathrm{DC}_\text{LLOA}\!\rightarrow\!\mathrm{AC}_{\text{SPF}}$ and
$\mathrm{DC}_{\text{BASE}}\!\rightarrow\!\mathrm{AC}_{\text{SPF}}$, dominate: 
for any error threshold, their curves lie above the single-slack ones, indicating 
fewer large generator-setpoint deviations. At $10^{-2}\,\text{p.u.}$, about 
$94.5\%$, $94\%$, $92.3\%$, and $91\%$ of generators lie below this error for 
the four pipelines, respectively. A notch near $10^{-1}\,\text{p.u.}$ on the 
$\mathrm{AC}_{\text{BASE}}$ curves reflects loss aggregation under a single slack, 
producing a cluster of similar deviations. The right tails confirm the same ordering: 
$\mathrm{AC}_{\text{SPF}}$ pipelines saturate near $1\,\text{p.u.}$, whereas 
$\mathrm{AC}_{\text{BASE}}$ ones require larger errors ($\sim\!1.2$–$1.3\,\text{p.u.}$) 
to reach $100\%$. These results underscore the advantage of distributed slack and 
advanced AC recovery in minimizing generator output errors and mitigating the 
limitations of single-slack aggregation.

\subsection{Bus-type Switching Effects on Voltage and Reactive Power}
\begin{figure*}[!t]
    \centering
    \includegraphics[width=1\textwidth]{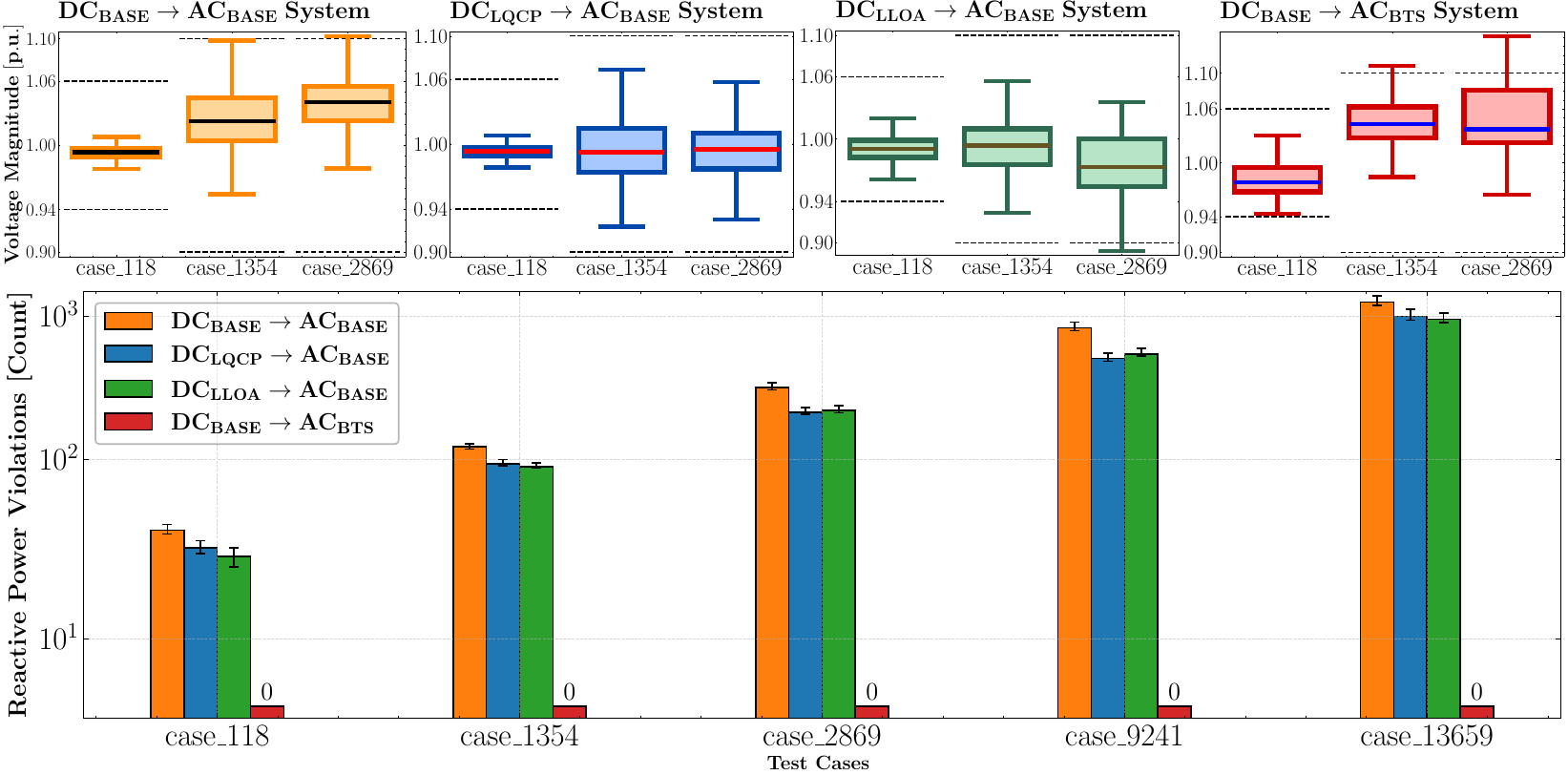}
    
    \caption{\text{Interplay between voltage magnitude and reactive power for
    various pipelines.} The plots are taken over load uncertainty with ${\sigma}=5\%$, for $\mathrm{1000}$ samples per case. The top box plots (mean with min–max  extremes) show the voltage distributions for three cases 
    (\texttt{case\_118}, \texttt{case\_1354}, and \texttt{case\_2869}). Dashed lines are voltage limits.
    The bottom plots show mean reactive power violations (with min–max error bars), with a y-axis log-scale, using:
    {\scriptsize $\mathrm{DC}_{\text{BASE}} \!\rightarrow\! \mathrm{AC}_{\text{BASE}}$} 
    (\textbf{\textcolor{orange}{orange}}), 
    {\scriptsize $\mathrm{DC}_{\text{LQCP}} \!\rightarrow\! \mathrm{AC}_{\text{BASE}}$} 
    (\textbf{\textcolor{blue}{blue}}), 
    {\scriptsize $\mathrm{DC}_{\text{LLOA}} \!\rightarrow\! \mathrm{AC}_{\text{BASE}}$} 
    (\textbf{\textcolor{ACgreen}{green}}), and 
    {\scriptsize $\mathrm{DC}_{\text{BASE}} \!\rightarrow\! \mathrm{AC}_{\text{BTS}}$} 
    (\textbf{\textcolor{red}{red}}).}

    \label{fig:Interplay}
        \vspace{-0.5em}
\end{figure*}

As discussed in Section~\ref{sec:qv}, reactive power and voltage are tightly coupled in ACPF. 
Fig.~\ref{fig:Interplay} shows this relationship across five test cases under load uncertainty $\sigma=5\%$. 
The upper box plots show voltage magnitudes with dashed limits, while the lower plots depict mean reactive violations. 
$\mathrm{AC}_{\text{BTS}}$ eliminates most reactive violations through dynamic PV/PQ switching, maintaining feasibility but causing larger voltage deviations—particularly in \texttt{case\_1354} and \texttt{case\_2869}, where voltages approach their upper limits. 
While voltage–reactive coupling allows PV/PQ switching to restore most voltage setpoints, near-limit voltages may demand reactive support beyond generator capability, leading to increased violations. 
$\mathrm{AC}_{\text{BASE}}$ enforces setpoints rigidly, yielding more reactive-limit breaches, whereas loss-aware $\mathrm{DC}_{\text{LQCP}}$ and $\mathrm{DC}_{\text{LLOA}}$ pre-account for losses and phase angles, reducing voltage deviations and violations. 
Among DC variants, $\mathrm{DC}_{\text{LQCP}}$ best maintained voltages within limits, even with $\mathrm{AC}_{\text{BASE}}$. 
Overall, $\mathrm{AC}_{\text{BTS}}$ demonstrates the benefits of reactive flexibility, while lossy DC pipelines help keep bus voltages near nominal values. 
These results highlight the voltage–reactive trade-off in AC feasibility recovery and the value of loss-aware setpoints for balanced operation. 
Voltage initializations from $\mathrm{AC}_{\text{BASE}}$ also improved convergence and reduced violations relative to a flat start.

\vspace{1em}

\section{Conclusion}

\label{sec:sect}
\hspace{0.5em} This paper investigates AC feasibility restoration pipelines for DCOPF dispatches. A comprehensive empirical study is conducted, applying various DCOPF and ACPF variants. The most effective and consistent pipeline for restoring AC feasibility from DC solutions is found to be $\mathrm{DC}_{\text{LQCP}}\!\rightarrow\! \mathrm{AC}_{\text{SPF}}$. The results show that integrating a structured ACPF—featuring distributed slack and reactive power limited generators—with loss-augmented DCOPF dispatches, can yield ACOPF feasible outcomes. This workflow reduces violations in active power, voltage, reactive power, and thermal limits while lowering the cost difference. For reference, applying the structured pipeline to the $13{,}659$-bus case achieved improvements of $93\%$ in cost difference, $75\%$ in mean absolute error, and an improved convergence rate compared to single-slack methods. The link between voltage and reactive power violations is also examined. Future work will focus on improving the pipeline’s computational efficiency, with a promising direction being the integration of parameterized DCOPF and structured ACPF in an end-to-end self-supervised learning framework to enhance AC feasibility, and scalability for DC-operated markets.

\begin{table*}[!t]
\centering
\renewcommand{\arraystretch}{1.0}
\caption{{\small\scshape Grouped vs. Sequential PV/PQ Switching: Total Iterations and Solving Time (Selected Cases)}}
\label{tab:grouped_switching_appendix} 
\scriptsize
\setlength{\tabcolsep}{3.5pt}

\sisetup{table-number-alignment=center}

\resizebox{\linewidth}{!}{
\begin{tabular}{
l
S[table-format=4.0] S[table-format=4.0] c
S[table-format=4.0] S[table-format=4.0] c
S[table-format=5.2] S[table-format=5.2] c
S[table-format=4.2]
}
\toprule
\textbf{Test case}
& \multicolumn{3}{c}{\textbf{Total Iterations: $\mathrm{AC}_{\text{BTS}}^{\star}$}}
& \multicolumn{3}{c}{\textbf{Total Iterations: $\mathrm{AC}_{\text{SPF}}^{\star}$}}
& \multicolumn{3}{c}{\textbf{Solving Time (s): $\mathrm{AC}_{\text{SPF}}^{\star}$}}
& \multicolumn{1}{c}{\textbf{Solving Time (s): ACOPF}} \\
\cmidrule(lr){2-4}\cmidrule(lr){5-7}\cmidrule(lr){8-10}\cmidrule(lr){11-11}
& {\scriptsize Sequential} & {\scriptsize Grouped} & {\scriptsize $\Delta$ (\%)}
& {\scriptsize Sequential} & {\scriptsize Grouped} & {\scriptsize $\Delta$ (\%)}
& {\scriptsize Sequential} & {\scriptsize Grouped} & {\scriptsize $\Delta$ (\%)}
& {\scriptsize } \\
\midrule
\texttt{case\_89}
& 42   & 4   & $\mathbf{{90.5}}$
& 12   & 5   & $\mathbf{{58.3}}$
& 0.23 & 0.26 & $-\mathrm{13.0}$
& 0.42 \\

\texttt{case\_118}
& 60   & 5   & $\mathbf{{91.7}}$
& 18   & 5   & $\mathbf{{72.2}}$
& 0.22 & 0.22 & $\mathrm{0.0}$
& 0.19 \\

\texttt{case\_1354}
& 357  & 11  & $\mathbf{{96.9}}$
& 152  & 10  & $\mathbf{{93.4}}$
& 64.10 & 6.05 & $\mathbf{{90.6}}$
& 2.78 \\

\texttt{case\_2869}
& 777  & 16  & $\mathbf{{97.9}}$
& 380  & 14  & $\mathbf{{96.3}}$
& 187.00 & 12.34 & $\mathbf{{93.4}}$
& 4.88 \\

\texttt{case\_9241}
& 2204 & 29  & $\mathbf{{98.7}}$
& 1356 & 22  & $\mathbf{{98.4}}$
& 6104.00 & 46.71 & $\mathbf{{99.2}}$
& 30.18 \\
\bottomrule
\multicolumn{11}{c}{\scriptsize $\star$ Sequential PV/PQ switching updates one violated generator per round; grouped switching updates multiple violated generators per round.}\\
\multicolumn{11}{c}{\scriptsize $\Delta(\%) = 100\cdot\left(1-\frac{\text{Grouped}}{\text{Sequential}}\right)$.}\\
\end{tabular}}
\end{table*}

\vspace{0.4em}

\IEEEtriggeratref{29}
\bibliographystyle{IEEEtran}
\bibliography{references}

\vspace{11em}

\section*{Appendix: \\Grouped PV/PQ Switching Runtime Comparison}
\addcontentsline{toc}{section}{Appendix: Grouped PV/PQ Switching Runtime Comparison}
\label{sec:appendix_grouped_switching}

This appendix explores the large runtime and iteration counts observed for sequential PV/PQ switching in large systems.
Table~\ref{tab:grouped_switching_appendix} compares sequential switching (updating one violated generator per switching round) against grouped switching (updating multiple violated generators per switching round) on selected cases.
For each test case, the first two column blocks report the \emph{total number of Newton iterations} required by $\mathrm{AC}_{\text{BTS}}$ and $\mathrm{AC}_{\text{SPF}}$ under sequential and grouped switching; thus, the reported totals aggregate the inner Newton iterations accumulated across all PV/PQ switching rounds (outer loop), and $\Delta(\%)$ gives the corresponding percent reduction.
The third block reports the resulting $\mathrm{AC}_{\text{SPF}}$ solving time under each strategy (again with $\Delta(\%)$), while the final column provides the ACOPF time for reference.

These results show that grouped switching substantially reduces the total Newton-iteration burden, with reductions typically exceeding $90\%$ for $\mathrm{AC}_{\text{BTS}}$ and ranging from roughly $60\%$ to approximately $98\%$ for $\mathrm{AC}_{\text{SPF}}$ across the selected cases.
This reduction in Newton-iteration workload directly translates into large runtime improvements in the cases where Table~\ref{tab:loss-functions2} exhibited high iteration counts and long solve times under sequential switching.
In particular, $\mathrm{AC}_{\text{SPF}}$ solve times improve by nearly two orders of magnitude in the most challenging cases (e.g., \texttt{case\_9241} drops from $6104$s to $46.71$s, a $\sim 130\times$ reduction).
Importantly, this improvement targets computational overhead from the switching procedure and therefore complements (rather than changes) the feasibility-restoration behavior discussed earlier: the cost difference and mean absolute error trends reported in Table~\ref{tab:loss-functions2} reflect the pipeline’s restoration quality, while Table~\ref{tab:grouped_switching_appendix} shows that the same restoration can be achieved much more efficiently by grouping violated buses per switching round.

\end{document}